\documentclass[amsmath, amssymb, twocolumn, superscriptaddress]{revtex4-2}
\usepackage{graphicx}
\usepackage{subcaption}
\usepackage{mathrsfs}
\usepackage{float}
\usepackage[colorlinks=true, citecolor=blue, linkcolor=magenta]{hyperref}%
\newtheorem{lemma}{Lemma}

\begin{document}

\title{Carter-Penrose diagrams for emergent spacetime in axisymmetrically accreting black hole systems}
\author{Susovan Maity}
\email{susovanmaity@hri.res.in}
\affiliation{Department of Physics, Harish-Chandra Research Institute, Chhatnag Road, Jhunsi, Allahabad 211 019, India}
\affiliation{Homi Bhabha National Institute, Training School Complex, Anushakti Nagar, Mumbai, 400094, Maharashtra, India}
\author{Md Arif Shaikh}
\email{arif.shaikh@icts.res.in}
\affiliation{International Centre for Theoretical Sciences, Tata Institute of Fundamental Research, Bangalore 560089, India}
\author{Pratik Tarafdar}
\email{pratikt@imsc.res.in}
\affiliation{The Institute of Mathematical Sciences, IV Cross Road, CIT Campus, Taramani, Chennai 600113, India}
\author{Tapas Kumar Das}
\email{tapas@hri.res.in}
\affiliation{Department of Physics, Harish-Chandra Research Institute, Chhatnag Road, Jhunsi, Allahabad 211 019, India}
\affiliation{Homi Bhabha National Institute, Training School Complex, Anushakti Nagar, Mumbai, 400094, Maharashtra, India}
\date{\today}	

\begin{abstract}
For general relativistic black hole accretion in the Kerr metric, we show that linear perturbation of the axially symmetric matter flow having certain geometrical configurations leads to the emergence of black hole like acoustic spacetime. Such an analogue spacetime is shown to be endowed with one white hole like sonic horizon flanked by two black hole like acoustic horizons. We construct the compactified causal structures, i.e., the Carter Penrose diagrams for such emergent spacetime to study the corresponding horizon effects. For the first time in literature, the Carter Penrose formalism is carried out for analogue spacetime embedded within a natural large scale fluid flow under the influence of strong gravity.
\end{abstract}

\maketitle

\section{Introduction}\label{introduction}
 Low angular momentum axisymmetric black hole accretion in the Kerr metric may manifest multi-transonic behaviour. In order to pass through more than one sonic point, accretion solutions must undergo a shock transition \cite{Fukue1983, Lu1985, Lu1986, Fukue1987, Nakayama1994, Yang1995, chakrabarti96mnras, Pariev1996, JFLu1997, Peitz1997, Takahashi, Barai2004, Nagakura2008, Nagakura2009, Das2012, Das2015, Tarafdar2015, Sukova2015JOP, Le2016, Sukova2017, Palit19mnras, Palit2020timevar, Palit2020AAS,  Tarafdar2021}. For such shocked accretion, subsonic flow of hydrodynamic matter starting from a large distance becomes supersonic after crossing a sonic point located at a relatively large distance from the black hole horizon. Such supersonic flow may then encounter a discontinuous stationary shock and becomes subsonic again. Shock induced subsonic flow then passes through another sonic point located at the close proximity of the horizon. A shocked multi-transonic accretion flow, thus, connects four different regions consisting of two subsonic regions and two supersonic regions from infinity to the innermost radius upto which the flow solutions can be extended.

Transonic accretion onto black hole can be posed as a two dimensional dynamical system problem as defined in dynamical systems theory. Using the structure of dynamical systems, it is possible to show that a stationary transonic black hole accretion solution may be realized as a critical solution on phase portrait characterized by the flow Mach number $M$ (defined by the ratio of the dynamical flow velocity and the characteristic sound speed) and the radial distance $r$ from the black hole horizon measured along the equatorial plane of the flow \cite{Ray_Arnab_PRE_2002, Ray_Arnab_MNRAS_2003, Ray2003, Ray2005_arxiv, Ray_2005, Ray_2007_CQG, bhattacharjee2007secular, bhattacharjee2009, ray2006, ray2007b, ray_khan}. A multi-transonic flow resembles a multi-critical phase orbit passing through two out of three critical points, where the two are saddle type critical points and the third one is a centre type critical point located in between aforementioned saddle points.

Whether any one to one mapping  between a critical point in the $M-r$ phase portrait and the corresponding sonic point of the flow exists, will actually depend on the geometrical configuration of the accretion flow. Axially symmetric hydrodynamic accretion in the Kerr metric may, in principle, be studied for following three different geometric configuration of the flow \cite{Chakrabarti2001, Bilic2014, Ho1999, Igumenshchev1999}:

i) Flow with constant thickness, where the flow thickness will not vary with the radial distance as measured along the equatorial plane of the disc. For any such flow considered in this paper, corresponding flow variables will be characterized by a subscript ${CH}$.

ii) Conical flow, for which the flow thickness is proportional to the radial distance $r$, i.e, the ratio of the flow thick and the radial distance is constant for all values of $r$. The flow geometry is then characterized by the solid angle $\Theta$ subtended by the disc at the horizon. This configuration is supposed to be the ideal-most one to portray a low angular momentum inviscid disc. For any such conical flow considered in this paper, corresponding flow variables will be characterized by a subscript ${CF}$.

iii) Flow in hydrostatic equilibrium along the vertical or the transverse (perpendicular to the equatorial plane) direction, for which the flow thickness is found to be a non linear function of the local radial distance. The expression for such flow thickness contains the expression of the characteristic radial sound speed, although the Euler equation is not formulated and solved along the vertical direction. For such flows considered in this paper, we will only use model prescribed by Novikov-Thorne which will be described later. Thus corresponding flow variables will be characterized by a subscript ${NT}$.

It has been observed that for general relativistic accretion flow in the Kerr metric with constant height flow as well as with conical flow, the critical points in the phase portrait and the corresponding sonic points coincide, i.e., the value of Mach number becomes unity at the critical points. For flow in hydrostatic equilibrium along the transverse direction, however, the Mach number becomes less than unity at the critical point for adiabatic equation of state of the accreting fluid. Hence, for the stationary definition of characteristic sound speed $c_s=\sqrt{{\gamma{p}}/\rho}$, the sonic point forms at a distance smaller compared to the location of the critical point, where $\gamma$ is the ratio of the specific heats at constant pressure and at constant volume, respectively ($\gamma=c_p/c_v$), $p$ and $\rho$ being the pressure and the mass density of accreting fluid.

The non isomorphism between the critical and the sonic points for flow in hydrostatic equilibrium along the transverse direction can be approached in two different ways. In conventional approach, one considers the expression of the characteristic sound speed, following the stationary definition of the same, and numerically integrate the flow equation inward (toward the black hole event horizon), starting from the critical point (using the critical point conditions) up to the radial distance where stationary Mach number becomes unity and designate that radial location to be the location of the sonic point (see, e.g., \cite{Das2012} for the detailed description of how one can locate the sonic point by numerically integrating the corresponding flow equations). Very recently, in an alternative approach, the dynamical definition of the sound speed has been found out for general relativistic accretion onto Schwarzschild black holes, by perturbing the flow equations using a space time dependent stability analysis. The expression of such effective sound speed differs from the stationary definition of $c_s$ and hence such dynamical sound speed has been dubbed as the `effective sound speed' $c_{\rm eff}$. If one considers the dynamically motivated effective sound speed instead of the stationary definition of sound speed, the critical and the sonic points coincide, and this isomorphism due to redefinition of sound speed is evident even from the stationary analysis of the accretion itself (see, e.g. \cite{Tarafdar2021} for further details).

Accreting black hole systems are quite versatile to study in the sense that not only such systems are investigated from the context of astrophysics as well as of the dynamical systems theory, but such systems can also be considered as realistic example of classical analogue gravity models. The analogue gravity phenomena is realized  by linearly perturbing a transonic fluid where a black hole like analogue spacetime structure with acoustic horizons emerges, where such analogue spacetime structure has a unique correpondence with a particular transonic flow line of the fluid considered \cite{Moncrief1980, Unruh81, Visser1998, Barcelo}. It has been established that by linearly perturbing the transonic accretion flow, a black hole like emergent acoustic space time can be produced within the accreting fluid \cite{Das2004, Dasgupta2005, Abraham2006, Das2007, Pu2012, Bilic2014, Tarafdar2015, Saha2016, Shaikh2017, Tarafdar2018, Shaikh2018a, Shaikh2018b, Shaikh2019}. An accreting black hole system is considered as a unique example of classical analogue gravity model since both kind of horizons, the analogue (acoustic, or sonic) as well as the gravitational one are present in the same system. The sonic points have usually been identified with  acoustic horizons, and for multi-transonic shocked accretion, the corresponding sonic geometry contains two black hole type acoustic horizons (which actually are two sonic points of the flow) and a white hole type acoustic horizon formed at the stationary shock location.

For constant height flow as well as for conical flow, the critical points are designated as the location of the acoustic horizons, since the critical points and the sonic points coincide for these flow geometries. For accretion in hydrostatic equilibrium along the transverse direction, however, there are ambiguities in determining the location of the acoustic horizons. If one considers the stationary definition of sound speed, then the sonic points are the acoustic horizons and not the critical points. On the other hand, if we chose the dynamical effective sound speed to be the speed of propagation of the acoustic perturbation traveling inside the linearly perturbed accretion flow, then the critical point itself becomes the sonic point and hence the acoustic horizons is supposed to form at the critical point. It is, however, to be noted that the definition of effective sound speed is essentially local, i.e., expression of such dynamical sound speed is defined based on its behaviour only at the corresponding critical points and such $c_{\rm eff}$ can only be defined globally, on every radial point, by analytical continuation imposed by the global redefinition. One thus needs to introduce a more concrete scheme for identifying the exact location of the acoustic horizon for analogue spacetime associated with accreting black hole  systems, as well as to understand the global structure of such non-trivial emergent spacetime.

The most common approach to the problem of analysing global structure of space-time consisting black holes and white holes, in the context of classical gravity, is to find a Kruskal like diagram for the corresponding space time metric. The coordinate transformations needed to write down the Kruskal like metric generally involves analytical continuation, i.e, the spacetime manifold represented by the complete range of the original coordinates becomes a subset of the spacetime manifold represented by the new Kruskal like coordinates. The uniformity and simple orientation of light cones make Kruskal like diagram a very suitable candidate for analysing global causal structure of such space time. But even in the context of classical general relativity, a space time consisting of black holes and white holes can be specified by a small number of parameters, thus making the space-time still much simpler than that is created by compact object. thus for this kind of simplistic spacetime one may consider the possibility of further analytical continuation, such that, the Kruskal like spacetime becomes a subset of much general manifold. In order to explore the possibility one must compactify the infinities of the Kruskal like coordinates into finite region. For example, just the introduction of spin parameter in black hole space time gives to rotating black hole spacetime, which can be described by Kerr metric. The Kruskal like diagram is obviously more interesting than Schwarzchild black holes, as the rotation of black hole becomes responsible for degenerate horizons. But compactifying the Kruskal like diagram for Kerr metric and further analytical continuation gives rise to much feature rich global structure involving infinite numbers of black holes and white holes in multiple universe connected in specific geometrical way. The best tool for the visualisation and interpretation for this kind of compactification is the Penrose-Carter diagram \cite{Carter66, Penrose1964, carter68, Boyer_lindquist_carter, walker_bock_diagram70}, not only for the Kerr or other kind of black hole space times, but also for cosmological space times. 

Motivated by the aforementioned discussions, in the present work we propose that construction of Carter-Penrose diagram will unambiguously identify the corresponding sonic horizons for accreting black hole systems, irrespective of the geometrical configuration of the flow. We first pick up accretion flow with certain geometry for which the critical points are the same as the sonic points. In this aspect the constant height flow and the conical flows are equivalent. We consider the conical flow since it portrays low angular momentum flow better in comparison to the constant height flow. We then find the critical points as well as the critical point conditions, and construct the corresponding phase portraits for such a flow. Then we linearly perturb the flow to obtain the analogue space time (thus the corresponding analogue metric) and construct the Carter-Penrose diagram using the metric elements of the acoustic metric to identify the acoustic black hole horizons at the critical points and the acoustic white hole horizon at the shock location. We perform such calculations for both adiabatic as well as for the isothermal flow.

We then consider accretion in hydrostatic equilibrium along the transverse direction for functional form of the local disc heights as introduced by Novikov and Thorne\cite{Novikov-Thorne1973} and mention the fact that the local disc height by Riffert and Herold\cite{Riffert-Herold1995} will have a similar treatment for reasons that will be clarified in subsequent sections. We perform the critical point analysis to find out the critical point conditions and show that the critical points are not the same as sonic points if one considers the stationary sound speed. We construct the corresponding phase portrait to identify the sonic points and the shock locations for the multi-transonic shocked flow. We then perturb the flow equations to construct the relativistic acoustic geometry and find out the expression for the effective sound speed as defined at the corresponding critical points. Finally we construct the corresponding Caretr Penrose diagram and establish that the acoustic horizons are formally located at the critical point of accretion flow in hydrostatic equilibrium along the vertical direction if one considers the dynamical definition of the effective sound speed. Since a Carter Penrose diagram can formally identify the location of the horizons, we ensure that the dynamical definition of effective sound speed is an important quantity when the critical points and the sonic points (sonic points defined in terms of the stationary sound speed) do not coincide for accretion onto a rotating black hole.

For the first time in literate, the Carter Penrose diagram has been constructed and used to study the analogue geometry embedded in transonic fluid for for any real physical system whose fluid properties are dependent on underlying physics. We thus use the Carter Penrose diagram to study the emergent space time embedded inside transonic accretion onto Kerr black holes.

\section{Governing Equations for Polytropic Flow and Choice of the Flow Thickness}
\noindent
We consider low angular momentum, inviscid, axially symmetric, irrotational accretion flow around a Kerr black hole for two different geometric configurations of the flow. The background metric and fluid equations are the same for both the flow geometries we are considering. The difference arises from the height prescriptions we consider where one disc has a conical cross-section, i.e, the height linearly rise with the radial distance and the other one is in hydrostatic equilibrium. Below the details of the models are specified. Then critical point analysis and causal structure analysis of emergent acoustic metrics are performed. In this work we will be working in the natural units of $G=1,c=1,M=1$ for convenience, where $G$ stands for the gravitational constant, $c$ is the speed of light and $M$ stands for the mass of the black hole in consideration.

\subsection{Background metric of Kerr black hole:}
\noindent
The background spacetime metric, as proposed by Boyer and Lindquist \cite{Boyer-Lindquist} at its equatorial plane can be expressed in terms of cylindrical coordinates and using the method of vertical averaging as described in \cite{Tarafdar_Deepika2019}, can be written as
\begin{equation}\label{background_metirc}
ds^2=-\frac{r^2 \Delta}{A}dt^2 + \frac{r^2}{\Delta}dr^2 + \frac{A}{r^2}(d\phi - \omega dt)^2 + dz^2.
\end{equation}
where
\begin{equation}\label{A-B-C}
 \Delta=r^2-2r+a^2,\quad A =r^2+ r^2 a^2+ 2ra^2, \quad \omega = \frac{2ar}{A}.
\end{equation}
Here $a = J/M$ is the Kerr parameter, Where $J$ and $M$ represents the total angular momentum  and the total mass of the rotating black hole, respectively. 

The physical horizon of Kerr black $r_+$ needed for analysing dynamics outside the physical black hole is  given by
\begin{equation}\label{r_plus}
 r_+ = 1+\sqrt{1-a^2}.
\end{equation}

\subsection{The Euler equation, continuity equation and the equation of state}
\noindent
The energy momentum tensor for a perfect fluid is given by
\begin{equation}\label{energy-mom-tensor}
T^{\mu\nu} = (p+\epsilon)v^\mu v^\nu + p g^{\mu\nu} ,
\end{equation}
where $ \rho $ is the rest-mass energy density of the fluid, so that $ \epsilon = \rho + \epsilon_{\rm thermal} $, and $p$ is the pressure of the fluid. $ v^\mu $ is the four-velocity with the normalization condition $ v^\mu v_\mu = -1 $.

 In the cylindrical Boyer-Lindquist frame, which is our choice for the coordinate system, the four velocity components can be expressed in terms of the advective velocity $u$, which is the three-component velocity in the co-rotating frame \cite{Gammie1998}. Now, the temporal component of four velocity $v_t$ in terms of $u$ is given by
 \begin{equation}\label{v_t}
 v_t = \sqrt{\frac{\Delta}{B(1-u^2)}}
 \end{equation}
 where $B = g_{\phi\phi} + 2\lambda g_{t\phi} -\lambda^2 g_{tt}$ and the specific angular momentum $\lambda$ is given by $\lambda = -v_{\phi}/v_t$. The radial component of the four velocity $v^r$ in terms of $u$ is given by
 \begin{equation}\label{v^r}
 v^r = \frac{u\sqrt{\Delta}}{r\sqrt{1-u^2}}
 \end{equation}
 
The fluid equations can not be analytically integrated if a barotropic fluid equation is considered. We first demonstrate the entire work for polytropic accretion and then summarize the entire work for isothermal accretion where complexities regarding non isomorphism of critical and sonic points do not arise. The equation of state for adiabatic flow is given by $ p = k\rho^\gamma $ where $\gamma$ is the polytropic index and $ k $ is constant. The sound speed for adiabatic flow (isoentropic flow) is given by
\begin{equation}\label{cs-ad}
c_{s}^2 = \left.\frac{\partial p}{\partial \epsilon}\right|_{\rm entropy} = \frac{\rho}{h}\frac{\partial h}{\partial \rho} ,
\end{equation}
where $ h $ is the enthalpy given by 
\begin{equation}\label{enthalpy}
h = \frac{p+\varepsilon}{\rho}
\end{equation}

The continuity equation and the Euler equation are given by, respectively,
\begin{equation}\label{continuity}
\nabla_\mu (\rho v^\mu) = 0
\end{equation}
and  
\begin{equation}\label{energy-mom-con}
\nabla_\mu T^{\mu\nu} = 0 .
\end{equation}
For adiabatic flow, the Euler equation simplifies as 
\begin{equation}\label{Euler}
v^\mu \nabla_\mu v^\nu +\frac{c_{s}^2}{\rho}(v^\mu v^\nu + g^{\mu\nu})\partial_\mu \rho = 0
\end{equation}
where expression of $ c_s^2 $ is given by equation (\ref{cs-ad}).

\subsection{Choice of disc heights}
As mentioned in section \ref{introduction}, simplistic accretion disc structure like conical disc, where $H(r)$ is a linear function of radial distance, has the property that the critical point turns out to be also the point where advective velocity is equal to the sound speed. Thus the critical points in this model turn out to be the sonic points also. Accretion disc with constant height also has the same property. But we will only consider conical flow as the representative of this kind of models with isomorphic critical points and sonic points as it portrays low angular moment accretion most accurately. The height of a conical disc as a function of radial distance is given by
\begin{equation}\label{CFheight}
  H_{CF}(r) = \Theta r
\end{equation}
where, as previously mentioned, $\Theta$ is the angular span of the conical structure.

There are three prescriptions for the height function in hydrostatic equilibrium, for none of which the critical points and sonic points coincide. The oldest, and most used expression for the disc thickness for accretion flow maintained in the hydrostatic equilibrium along the vertical direction, was provided by Novikov \& Thorne \cite{Novikov-Thorne1973} as
\begin{widetext}
\begin{equation}\label{NTheight}
H_{NT} (r)= \left(\frac{p}{\rho}\right)^\frac{1}{2} \frac{r^3 + a^2 r + 2a^2 }{r^{\frac{3}{2}} + a} \sqrt{\frac{r^6 - 3r^5 + 2ar^{\frac{9}{2}}}{(r^2 -2r +a^2)(r^4 + 4a^2 r^2 -4a^2 r + 3a^4 )}}
\end{equation}
\end{widetext}
It is to be noted that accretion flow described by the above disc thickness can not be extended up to $r_+$. The flow will be truncated at a distance $r_T$, where
\begin{equation}\label{r_truncated}
\left( r_T \right)^{\frac{1}{2}} (r_T -3) = 2a 
\end{equation}
which is outside $r_+$. In reality of course the flow will exist up to $r_+$ but no stationary integral flow solutions can be constructed up to the close proximity of $r_+$ for accretion flow described by the disc height prescribed by Novikov and Thorne.

Riffert and Herold \cite{Riffert-Herold1995} provided an expression of disc thickness by modifying the gravity-pressure balance condition of the treatment done in Novikov \& Thorne. In the work of Novikov and Thorne, the vertical component of gravity was replaced by $zR^{z}_{0z0}$ in the vertical component of Newtonian gravity pressure balance equation. Riffert and Herold on the other hand used the Euler equation directly to find out the gravity pressure balance equation and the height expression as formulated by them is given by

\begin{equation}\label{RHheight}
H_{RH} (r) = \left(\frac{p}{\rho}\right)^\frac{1}{2} \sqrt{\frac{r^5 - 3r^4 + 2ar^{\frac{7}{2}}}{r^2 -4ar^{\frac{1}{2}} + 3ar^2}}
\end{equation}

In this case also the flow can only be extended upto $r_T$ as given by (\ref{r_truncated}). We see that both the disc heights can be expressed in the form by $H(r)= (p/ \rho)^{1/2}f(r,a)$. The prefactor in this expression is related to sound speed and thus these two height expressions are dependent on the flow variable itself. As will be seen later, this particular form of both the disc heights given by eq. (\ref{NTheight}) and eq. (\ref{RHheight}) is responsible for the fact that critical points do not coincide with the transonic points. This particular general form of the height expressions for both these models are responsible for the fact that they also have same Mach number at the critical points. Thus choosing any one of these models will be sufficient to demonstrate this kind of non isomorphism of critical points and sonic points.

More recently Abramowicz, Lanza and Percival \cite{Abramowicz1996ap} introduced an expression for the disc thickness, given by
\begin{equation}\label{ALPheight}
H_{ALP} (r) = \left(\frac{p}{\rho}\right)^\frac{1}{2} \sqrt{\frac{2r^4}{v_\phi^2 - a^2 (v_t - 1)}} ,
\end{equation}
where $v_\phi$ and $v_t$ are the azimuthal and time component of the four velocity of accreting fluid.For this height function, the steady state accretion solutions can be obtained upto $r_+$. It is, however to be noted that by linearly perturbing flow equation for flow thickness (\ref{ALPheight}), acoustic metric could not be constructed as of now. Thus we will not be using this height expression for stationary analysis as it can not be used for later analysis involving acoustic metric.

In our present work, we thus use the height function due to Novikov \& Thorne only as a representative of models where critical points are not transonic points.

Although the complete specification of the accretion model we will be working with hass been presented, we must clarify that, certain terminology regarding the velocities and other quantities defined above will be used throughout this work. The reason is that, a perturbation analysis will be used to study dynamics of first order perturbations, which will be performed on steady state flow. Thus we must use a nomenclature to distinguish between the steady state and first order flow. We will frequently denote $v^{\mu}_{0}$, $u_0$ as four-velocity and advective velocity corresponding to the steady-state flow. In general, we will use the subscript zero to denote the value of any physical variable corresponding to the stationary solutions of the steady flow, e.g., $p_0 , \rho_0$ etc. We will define the first order perturbation of any physical variable to have a subscript of one where the perturbation analysis will be carried out. No use of subscript on any variable in some equation denotes that the equation is valid in dynamical case and the variable is the sum of the steady state part and the dynamical first order part. 

\section{Description of Multitransonic Adiabatic Flow as Dynamical System Problem}\label{Sec:Crit-points}
In order to represent the problem of stationary accretion of ideal fluid following adiabatic equation of state around a rotating black hole, one must find the specific form the governing fluid equations as specified in eq. (\ref{continuity}) and eq. (\ref{Euler}) for the two height functions as specified in eq. (\ref{CFheight}) eq. (\ref{NTheight}). It will be shown in this section that solving the problem of stationary accretion, i,e, describing the dynamics of a compressible astrophysical fluid essentially boils down to solving a set of differential equations involving the derivative of advective velocity $du_0 / dr$ and derivative of $dc_{s_0}$. The problem can also be dealt by solving one differential equation involving the expression of $du_0 / dr$ and an algebraic equation simultaneously. This later approach will be followed in this work. Thus the plan of work  in this section will be as follows.

We will first integrate the general forms of the dynamical equation and obtain two integrals of motion as one of the expressions involving the specific energy, denoted by $\mathcal{E}_0$, will be used as the aforementioned algebraic equation. Next we derive the specific form of the derivative of advective velocity $du_0 / dr$. Here we will see how the analytical structure of accreting fluid system can be formulated as a problem of dynamical system. The initial condition relating sound speed and advective velocity of the fluid at the critical points will also be derived for both kind of accretion flows having a conical disc height and disc height as proposed by Novikov and Thorne. From the critical conditions, the isomorphism of critical and sonic points for accretion flow with conical disc and non isomorphism of critical and sonic points for accretion flow with disc height as proposed by Novikov and Thorne will be pointed out. In the next part it will be demonstrated how multi transonic solutions of the accretion can be constructed from critical solutions of the dynamical system of equations for specific subset of the parameters of the problem. Once one obtains the multi transonic solution, the flow line is determined and we will proceed for the perturbation analysis in the next section.

\subsection{The first integrals of motion}
In order to establish the problem of accretion as a dynamical system, two first integrals of motion for the stationary flow are needed. We first deal with the Euler equation as the height function does not enter in the equation. Then we differentiate between the conical flow and Novikov-Thorne type disc when we deal with the continuity equation.

The conseravtion of the temporal component of the simplified version of Euler equation given by eq. (\ref{Euler}) leads to the constancy of specific energy of the accreting fluid $\mathcal{E}_0$, given by
\begin{equation}\label{bernoulli_constant_ad}
\mathcal{E}_0 = \frac{\gamma - 1}{\gamma - 1 - c_{s0}^2}\sqrt{\frac{\Delta}{B(1-u_0^2)}}
\end{equation}

The second integral of motion obtained by integrating continuity equation is the mass accretion rate $\dot{M}$, although we will later use $\Psi$ instead of $\dot{M}$ for simplification of notation. $\dot{M}$ can be expressed as 
\begin{equation}\label{mass_accretion_rate}
\dot{M} = 4\pi H(r)r\rho \frac{\sqrt{\Delta}u_0}{r\sqrt{1-u_0^2}}.
\end{equation}
The entropy accretion rate $\dot{\Xi}$ is proportional to $\dot{M}$ and can be expressed as
\begin{equation}\label{entropy_accretion_rate}
{\dot \Xi} = \left( \frac{1}{\gamma} \right)^{\left( \frac{1}{\gamma-1} \right)} 4\pi \Delta^{\frac{1}{2}} c_{s}^{\left( \frac{2}{\gamma - 1}\right) } \frac{u}{\sqrt{1-u^2}}\left[\frac{(\gamma -1)}{\gamma -(1+c^{2}_{s})}\right]^{\left( \frac{1}{\gamma -1} \right) } H(r)
\end{equation}

We thus have two primary first integrals of motion along the streamline -- the specific energy of the flow ${\cal E}$ and the mass accretion rate ${\dot M}$. Even in the absence of creation or annihilation of matter, the entropy accretion rate ${\dot \Xi}$ is not a generic first integral of motion. As the expression for ${\dot \Xi}$ contains the quantity $K{\equiv}p/{\rho}^{\gamma}$, which is a measure of the specific entropy of the flow, the entropy accretion rate ${\dot \Xi}$
remains constant throughout the flow only if the entropy per particle remains locally invariant. This condition may be violated if the accretion is accompanied by a shock and as we will see in the next section, multi transonic solution will have a shock present in the flow line. However, $\dot{\Xi}$ is an important quantity needed to distinguish between flows having different topological structure in phase portraits for different values of parameters. We do not study the parametric variation of flows as we only focus on accretion flow where shock will be present. Still, $\dot{\Xi}$ is required in order to find the expression of $du_0 / dr$. 

\subsection{Velocity gradient and derivation of critical conditions}
One obtains two linear equations involving the derivative of the advective velocity $du_0 / dr$ and the derivative of sound speed $dc_{s0} / dr$ by taking derivatives of the two constants of integration. If the expression of $dc_{s0} / dr$ from one equation is substituted in the other equation, then one gets the expression of the derivative of advective velocity. While taking the derivative of eq. (\ref{entropy_accretion_rate}) the two height expressions given by eq. (\ref{CFheight}) and eq. (\ref{NTheight}) are used separately.

The expression for derivative of advective velocity $u_0$ corresponding to the flow with conical height function turns out to be (see \cite{Tarafdar_Deepika2019})
\begin{widetext}
\begin{equation}\label{dudr_CF}
\frac{du_0}{dr}\Big|_{CF} = \frac{u_0(1-u_0^2)\left[c_{s0}^2\frac{2r^2 - 3r + a^2}{\Delta r}+\frac{1}{2}(\frac{B'}{B}-\frac{\Delta'}{\Delta})\right]}{u_0^2- c_{s0^2}}=\frac{N_{CF}}{D_{CF}}.
\end{equation}
\end{widetext}

 The expression for derivative of advective velocity $u_0$ corresponding to the flow with height function as prescribed by Novikov and Thorne turns out to be (see \cite{Tarafdar2021})
\begin{widetext}
\begin{equation}\label{dudr_NT}
\frac{du_0}{dr}\Big|_{NT} = \frac{u_0(1-u_0^2)\left[\frac{2}{\gamma +1}c_{s0}^2(\frac{\Delta'}{2\Delta}+\frac{f'}{f})+\frac{1}{2}(\frac{B'}{B}-\frac{\Delta'}{\Delta})\right]}{u_0^2-\frac{c_{s0^2}}{(\frac{\gamma +1}{2})}}=\frac{N_{NT}}{D_{NT}}
\end{equation}
\end{widetext}

In both the equations mentioned above, the numerator is denoted as $N$ and denominator as $D$. One can thus define some parameter $\tau$ such that $du_0/d \tau = N, dr/ d\tau = D$. In this way the problem of finding all the time independent fluid profile of accretion can be posed as a problem of dynamical systems. Now we will use the tools of non linear dynamics to find out the critical points and then later to draw the critical flows in phase portraits.

The critical points can be obtained by setting $du_0/d \tau = 0, dr/ d\tau = 0$ simultaneously. Thus the critical point conditions turn out to be $N = 0$ and $ D = 0$.

Thus for conical flow, the condition $D=0$ at critical point yields
\begin{equation}\label{crit_CF}
u_0^2\big|_{r = r_c} =c_{s0}^2\big|_{r = r_c}
\end{equation}
whereas for flow with height function as prescribed by Novikov and Thorne, the condition $D=0$ turns out to be
\begin{equation}\label{crit_NT}
u_0^2\big|_{r = r_c} = \frac{c_{s0}^2 \big|_{r = r_c}} {(\frac{\gamma +1}{2})}
\end{equation}
or for later convenience we can write
\begin{equation}\label{beta_defn}
u_0^2\big|_{r = r_c} = \frac{c_{s0}^2\big|_{r = r_c}}{1+\beta},\quad {\rm where}\quad \beta = \frac{\gamma-1}{2}. 
\end{equation}
The key point we get from eq. (\ref{crit_CF})) and eq. (\ref{crit_NT})) is that the Mach number, i.e, the ratio of advective velocity and sound speed, is $1$ at the critical points for conical flow. But for disc height as prescribed by Novikov and Thorne, the Mach number is $\sqrt{1/1+\beta}$ at critical points, which is always less than unity for $\gamma > 1$. Thus the  critical points and the sonic points are isomorphic in the case of Conical Flow, but for disc height described by Novikov and Thorne, the sonic and critical points are not isomorphic.

The value of sound speed at critical points can be obtained from the other critical condition $N = 0$ and substituting the value of sound speed and corresponding value of advective velocity in eq. (\ref{bernoulli_constant_ad}), one obtains the radii of the critical points. The value of the derivative of the advective velocity at the critical points can also be obtained and thus along with the locations of the critical points, the complete set of initial conditions needed to integrate eq. (\ref{dudr_CF}) and eq. (\ref{dudr_CF}) are obtained. At this point one needs to specify the parameters of the problem so that the phase portraits can be plotted by solving the expressions for the derivatives of advective velocity.

\subsection{Nature of critical points and phase portrait for multi transonic accretion}
The detailed method of obtaining a multitransonic flow line in the phase portrait from the initial critical conditions corresponding to conical flow has been described in \cite{Tarafdar2018} and the same method for disc height as prescribed by Novikov and Thorne has been described in \cite{Tarafdar2021}. But in order to perform perturbation analysis on a certain flow line, we must first describe the nature of the critical points and the solutions passing through them, namely the critical flow lines. Finally we choose a particular flow line consisting of the critical flow lines such that  multi transonicity is achieved.

To numerically obtain phase portraits eq. (\ref{dudr_CF}) and eq. (\ref{dudr_NT}) has to be integrated and one needs to specify the parameters needed to solve this problem which are present in the equations itself and in the initial conditions. There are four parameters given by $\mathcal{E}_0, \lambda, \gamma$ and $a$ for both kind of flows with the conical height function and the height function as prescribed by Novikov and Thorne. The topology of the flow is dependent on the values of parameters. As previously mentioned, we want to focus on the characteristics of multitransonic accretion in this work. For the dynamic system corresponding to the models of accretion we are concerned with, there exists only one sonic point corresponding to each saddle type critical point. A physically acceptable transonic solution can be constructed only through a saddle-type critical point, and not through a centre type critical point. Hence the concept of a sonic point corresponding to a centre type critical point is meaningless. Thus in order to choose a multi transonic accretion flow, we must choose the parameters such that the dynamical system corresponding to the flow has multiple critical points and the critical flows correspond to accretion and not wind.

\begin{figure*}[!t]
\begin{subfigure}{.5\textwidth}
    \centering
    \includegraphics[width=.8\columnwidth]{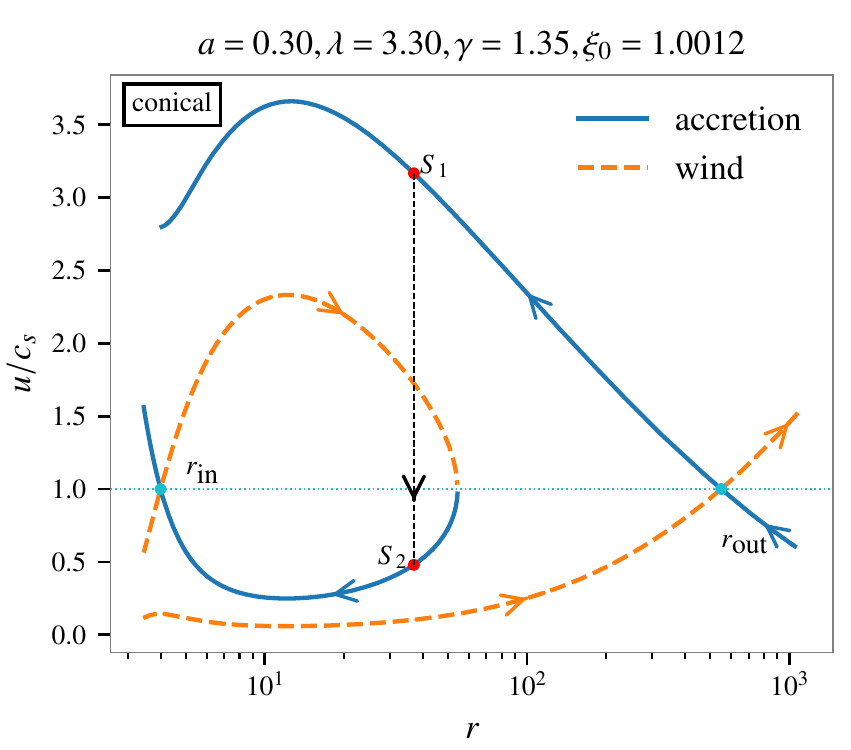}  
    \caption{Phase portrait for adiabatic flow with conical disc height}
    \label{fig:conical_ad}
\end{subfigure}%
\begin{subfigure}{.5\textwidth}
    \centering
    \includegraphics[width=.8\columnwidth]{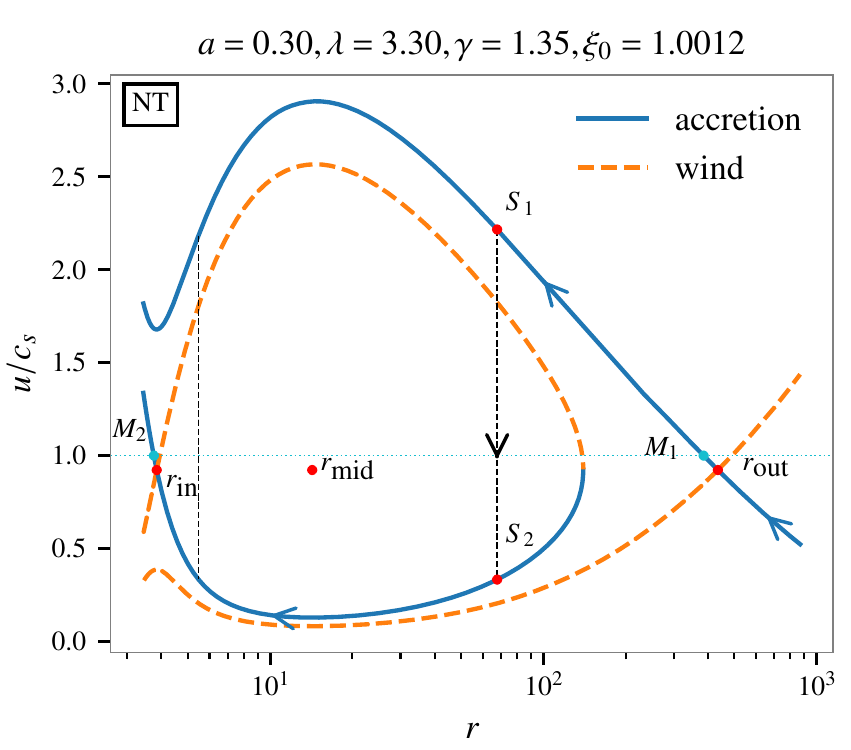}  
    \caption{Phase portrait for adiabatic flow with height expression formulated by Novikov and Thorne}
    \label{fig:NT_ad}
\end{subfigure}

\caption{Both phase portraits have been drawn for adiabatic accretion with the set of parameter values $\mathcal{E}_0=1.0012$, $\lambda=3.30$ and $\gamma=1.35$. For both pictures the blue solid lines corresponds to accretion branch whereas the orange dashed lines correspond to wind branch. The innermost critical point is denoted as $r_{\rm in}$ and outermost critical point is denoted as $r_{\rm out}$. $S_1$ corresponds to the point in phase portrait in accretion branch through outer critical point where shock may occur and $S_2$ corresponds to the point where shock occurs in accretion branch through inner critical point. The same radial distance of $S_1$ and $S_2$ corresponds to the fact that the shock is infinitesimally thin. The dotted black line joining $S_1$ and $S_2$ corresponds to the discontinuous jump in the shock location. In fig (\ref{fig:NT_ad}) the phase portrait of flow with height as prescribed by Novikov and Thorne, the Mach number at critical points are not unity and thus the critical point and sonic points do not coincide for this height prescription. The outer sonic point and the inner sonic point in the phase portrait corresponding to the flow with height function as prescribed by Novikov and Thorne are respectively $M_1$ and $M_2$}
 \label{fig:phase-portrait}
\end{figure*}

The numerical plots of the critical flows in the phase portraits for both kind of flows with conical height function and height function as prescribed by Novikov and Thorne are presented in in fig. (\ref{fig:phase-portrait}), such that the chosen parameters correspond to multicritical accretion for both the flows.

In both the figures portrayed in fig. (\ref{fig:phase-portrait}), $r_{\rm out}$ and $r_{\rm in}$ corresponds to the outer and inner critical point respectively, both of which are saddle type critical points. The critical point which resides in the middle of the outermost critical point located at $r_{\rm out}$ and innermost critical point $r_{\rm in}$, is centre type in nature, although not shown explicitly in the plot of critical flow lines as the point plays no role in a multi transonic flow line.
 
In fig. (\ref{fig:phase-portrait}), the parameters of the problems are chosen such that the phase portrait corresponds to accretion flow with multiple critical points. But the presence of multiple critical points does not itself ensure the presence of multi transonic flow. From the figures it is clear that a continuous critical flow through any one of this two saddle type critical points in the phase portrait will not pass through the other saddle type critical point. One can only choose a multi transonic flow consisting of the critical solutions if there is a mechanism that allows a discontinuous jump from the supersonic critical flow line through the outer critical point  $r_{\rm out}$  to the subsonic critical flow line through the innermost critical point $r_{\rm out}$. This discontinuity in an accretion flow  can be physically modeled by a Rankine-Hugoniot type infinitesimally thin shock. There is a subset of parameters within the parameter space corresponding to multi transonic accretion, where shock may occur. In the plot of the critical solutions in the phase portrait, the parametrs are chosen for both the flow in such a way that such a shock occurs. The fluid elements in the disc accretion flows in hydrostatic equilibrium irrespective of the height function of the disc, start far away from the accretor, in this case, the black hole, along the critical flow line and move through the outer critical point $r_{\rm out}$, continue along the flow line (the blue solid line passing through $r_{\rm out}$ in the online version of the figure) and will be supersonic till the flow variables and thermodynamic variables make a discontinuous jump to another branch (the green dotted line in the online version of the figure) as a consequence of shock. The discontinuous jump occurs at $S_1$ where the flow passing through the outer critical point is supersonic and reaches $S_2$ on the flow line passing through the inner critical point, where the flow becomes subsonic again. Then the fluid element eventually goes through $r_{\rm in}$. This discontinuity is strictly a discontinuity of the values of the fluid variables and not a discontinuity present in the physical flow.  In this way the presence of shock makes the fluid element move through a multi transonic flow line. Thus we not only choose the parameters to allow multiple critical points but also to allow multi-transonic accretion.

By following the aforementioned scheme, we are able to choose a multitransonic accretion flow line for each of the flow corresponding to the conical height function and height function as prescribed by Novikov and Thorne. The major difference between the flowlines of the two models is that whereas the flow achieves transonicity at $r_{\rm out}$  and again at $r_{\rm in}$ for conical flow, the flow for disc height as prescribed by Novikov and Thorne does not, as the Mach number is lesser than one at $r_{\rm out}$ and $r_{\rm in}$. For flow with disc height as prescribed by Novikov and Thorne, a fluid element of the flow becomes subsonic to supersonic for the first time at $M_1$ and  after it becomes supersonic to subsonic at the shock location, it becomes supersonic for the second time at $M_2$ as shown in fig. (\ref{fig:NT_ad}).

Before concluding the discussions about the stationary solutions of accretion flow, we note that, we have so far established the non homomorphism of critical and sonic points for flow with disc height as prescribed by Novikov and Thorne. One can simply assert that the fact that the critical point and sonic point do not coincide, is just a feature of the height function as prescribed by Novikov and Thorne. This feature does not necessarily makes us question, how correct is the definition of sound speed is in a dynamical context. To answer this question one may try to perturb the dynamical equations governing the flow and try to find out how linear perturbation behaves as sound speed at its core is the speed of propagation of first order perturbation in a medium. The systematic analysis of perturbations of fluid equations leads to treating certain linearly perturbed fluid variable from the context of analogue or emergent gravity. The perturbation analysis has been performed in the next section and the connection with analogue gravity is established.

\section{Derivation of Acoustic Metric from Linear Perturbation of Fluid Equations}
The non homomorphism of critical points and sonic points established in the previous section does not necessitate one to redefine sound speed as the stationary picture does not suggest how the acoustic perturbation propagates in the accreting medium. The motivation of this work will thus only be clear when one performs a perturbation analysis of the full spacetime dependent fluid equations. In this section we perturb the dynamical equations describing adiabatic accretion flow in the background metric of rotating black hole for conical disc height as well as disc height as prescribed by Novikov and Thorne. The perturbations yield dynamical equations governing the first order perturbation of certain flow variable. Then acoustic metric is written by comparing the aforementioned governing equations with the equation of a massless scalar field in curved spacetime. The basic methodology of analogue gravity is based on identifying the similar forms of the governing equation of first order perturbation of certain fluid variable and the massless scalar field equation in curved spacetime. The analogy is subject to certain conditions on the flow which are already fulfilled by our choice of ideal inviscid fluid. Thus for acoustic metric describing the perturbation in the accretion flow with height function as prescribed by Novikov and Thorne, we show that, the the acoustic black hole horizons are located at the critical points. To conclude this section we define an effective sound speed, for which the sonic points turn out to be critical points.

As our work concerns two disc models, we must differentiate the two emergent metrices as well. Although that will be done by the end of this section, it must be mentioned how the differences in the functional forms of conical disc height and disc height as prescribed by Novikov and Thorne manifest itself in the perturbation analysis. If we compare eq. (\ref{CFheight}) describing the height function for conical flow and eq. (\ref{NTheight}) describing the height function formulated by Novikov and Thorne, we see that the perturbation of flow variable will have no consequence on the perturbation of conical flow as it is only a function of radial distance and on the other hand it will be directly manifested in the perturbation of flow with disc height prescribed by Novikov and Thorne as it directly depends on the sound speed, a flow variable, in case of adiabatic flow. Thus we will differentiate between the two models where perturbation of height is introduced in this section and the parameter distinguishing the two flows will turn out to be $\beta$ as defined in eq. (\ref{beta_defn}) corresponding to only flow with height function as prescribed by Novikov and Thorne. We will use the notation $\beta$ for conical flow also and fix its value to zero so that the equations describing the perturbations have the same structural form. When we write down the acoustic metric we will use $\beta$ only for flows with height function as prescribed by Novikov and Thorne and remove $\beta$ from the expression of acoustic metric for conical flow by setting it zero.

Now the perturbation on the stationary flow is done by following standard linear perturbation analysis \cite{deepika2015,deepika2017,Shaikh2017,Shaikh2018a} where acoustic space-time metric for conical flow was derived. Time-dependent accretion variables, like the components of four velocity and pressure are written as small time-dependent  linear perturbations added their stationary values. Thus we can write,
\begin{equation}\label{perturbations}
\begin{aligned}
& v^t(r,t) = v^t_0(r)+{v^t_1}(r,t)\\
& v^r(r,t) = v^r_0(r)+{v^r_1}(r,t)\\
& \rho(r,t) = \rho_0(r)+\rho_1(r,t)
\end{aligned}
\end{equation}
where the subscript `1' denotes the first order small perturbation of some variable about the stationary value denoted by subscript `0'. The second constant of integration from continuity equation $\dot{\mathcal{M}}$ or $\Psi$ has the form (see \cite{Tarafdar_Deepika2019}, \cite{Tarafdar2021}) 
\begin{equation}\label{Sationary-mass-acc-rate}
\Psi = 4\pi \sqrt{-g}  \rho(r,t) v^r(r,t) H_\theta
\end{equation}
which is the stationary mass accretion rate of the accretion flow. Thus
\begin{equation}\label{psi-perturbation}
\Psi(r,t) = \Psi_0 + \Psi_1(r,t) 
\end{equation}
Where $ \Psi_0 $ is the stationary mass accretion rate defined in equation (\ref{Sationary-mass-acc-rate}). The constants can be absorbed in the definition without any loss of generality. Using the eq. (\ref{perturbations}) we get 
\begin{equation}\label{Psi1}
\Psi_1 = \sqrt{-g}[\rho_1 v_0^r H_{\theta0}+\rho_0 v^r_1H_{\theta 0}+\rho_0 v_0^rH_{\theta 1}]
\end{equation}
The last term in $ \Psi_1 $ consists of a term with the perturbation of angular height function $ H_\theta $. We recall $ H(r) $ as $ H_{\theta} = H(r)/r$. For now, general height function $H(r)$ is used whereas later on we distinguish between conical flow and disc height proposed by Novikov and Thorne.

For adiabatic flow (\ref{enthalpy}) can be rewritten as 
\begin{equation}
h=1+\frac{\gamma}{\gamma-1}\frac{p}{\rho}
\end{equation}
where the perturbed quantity $ h_1 $ can be written as
\begin{equation}\label{h1}
h_1 = \frac{h_0 c_{s0}^2}{\rho_0}\rho_1
\end{equation}
For adiabatic flow the irrotationality condition is \cite{Tarafdar2021}
\begin{equation}\label{irrot}
\partial_\mu(hv_\nu)-\partial_\nu(hv_\mu) = 0
\end{equation}
Now, using eq. (\ref{irrot}), the normalization condition $v^\mu v_\mu=-1$ and the axial symmetry of the flow we obtain quantities needed for further perturbation.
From irrotaionality condition eq. (\ref{irrot}) with $\mu=t$ and $\nu=\phi$ and with axial symmetry we have
\begin{equation}\label{irrotaionality_t_phi}
\partial_t(h v_\phi)=0,
\end{equation}
and for $\mu=r$ and $\nu=\phi$ and using axial symmetry, we have
\begin{equation}\label{irrotationality_r_phi}
\partial_r(h v_\phi)=0.
\end{equation}
So $h v_\phi$ is a constant of motion and eq. (\ref{irrotaionality_t_phi}) gives
\begin{equation}\label{del_t_v_phi}
\partial_t v_\phi=-\frac{v_\phi c_s^2}{\rho}\partial_t \rho.
\end{equation}
Using  $v_\phi=g_{\phi\phi}v^\phi+g_{\phi t}v^t$ in the previous equation gives
\begin{equation}\label{del_t_v_up_phi}
\partial_t v^\phi=-\frac{g_{\phi t}}{g_{\phi\phi}}\partial_t v^t-\frac{v_\phi c_s^2}{g_{\phi\phi}\rho}\partial_t \rho.
\end{equation}
The normalization condition of four velocity $v^\mu v_\mu=-1$ in this case can be written as
\begin{equation}\label{Normalization_condition}
g_{tt}(v^t)^2=1+g_{rr}(v^r)^2+g_{\phi\phi}(v^\phi)^2+2g_{\phi t}v^\phi v^t.
\end{equation}
The time derivative of this equation is 
\begin{equation}\label{del_t_v_up_t_1}
\partial_t v^t=\alpha_1\partial_t v^r+\alpha_2\partial_t v^\phi
\end{equation}
where $\alpha_1=-{v_r}/{v_t}$, $\alpha_2=-{v_\phi}/{v_t}$ and $v_t = -g_{tt}v^t+g_{\phi t}v^\phi$. Replacing $\partial_t v^\phi$ in eq. (\ref{del_t_v_up_t_1}) using eq. (\ref{del_t_v_up_phi}) we get

\begin{equation}\label{del_t_v_up_t_2}
\partial_t v^t=\left( \frac{-\alpha_2v_\phi c_s^2/(\rho g_{\phi\phi})}{1+\alpha_2 g_{\phi t}/g_{\phi\phi}}\right)\partial_t \rho+\left(\frac{\alpha_1}{1+\alpha_2 g_{\phi t}/g_{\phi\phi}} \right)\partial_t v^r
\end{equation}

Using eq. (\ref{perturbations}) in eq. (\ref{del_t_v_up_t_2}) and collecting the linear perturbation part we get
\begin{equation}\label{del_t_pert_v_up_t}
\partial_t v_1^t=\eta_1\partial_t \rho_1+\eta_2\partial_t v^r_1
\end{equation}
where
\begin{widetext}
\begin{equation}\label{eta_1_eta_2_and_Lambda}
\eta_1=-\frac{c_{s0}^2}{\Lambda v^t_0\rho_0}[\Lambda (v^t_0)^2-1-g_{rr}(v^r_0)^2],\quad \eta_2=\frac{g_{rr}v^r_0}{\Lambda v^t_0}\quad {\rm and}\quad \Lambda=g_{tt}+\frac{g_{\phi t}^2}{g_{\phi\phi}}
\end{equation}	
\end{widetext}
Now we perturb height function and differentiate between the two models we consider. For conical flow, $H_{CF} = \Theta r$. Thus
\begin{equation}
(H_{\theta 1})_{CF} = 0
\end{equation}
For NT, $(H_{\theta1})_{NT}$ can be written as 
\begin{equation}\label{h_theta1}
\frac{(H_{\theta1})_{NT}}{(H_{\theta 0})_{NT}} = \left(\frac{\gamma -1}{2} \right)\frac{\rho_1}{\rho_0} = \beta_{NT} \frac{\rho_1}{\rho_0}
\end{equation}
where $\beta_{NT} = \gamma -1/2$. Here we see that the whole perturbation analysis can be generalized if we define $\beta_{CF} = 0$  for conical flow and continue the analysis with $\beta$ in general. In the end we can again put these two different values and obtain different results for the two models.

The continuity equation takes the form 
\begin{equation}\label{conserve}
\partial_t(\sqrt{-g}\rho v^t H_\theta)+\partial_r(\sqrt{-g}\rho v^r H_\theta) = 0.
\end{equation}
Using eq. (\ref{perturbations}) and  (\ref{psi-perturbation}) in the previous equation and using eq. (\ref{del_t_pert_v_up_t}) and (\ref{h_theta1}) and replacing them in (\ref{conserve}) yields
\begin{equation}\label{del_r_psi_1}
-\dfrac{\partial_r\Psi_1}{\Psi_0} = \dfrac{\eta_2}{v_0^r}\partial_t v^r_1+\dfrac{v_0^t}{v_0^r \rho_0 }\left[ 1+\beta +\frac{\eta_1 \rho_0}{v_0^t}\right]\partial_t \rho_1,
\end{equation}
and
\begin{equation}\label{del_t_psi_1}
\dfrac{\partial_t\Psi_1}{\Psi_0} = \dfrac{1}{v_0^r}\partial_t v^r_1+ \dfrac{1+\beta}{\rho_0}\partial_t \rho_1.
\end{equation}
Using the two equations given by eq. (\ref{del_r_psi_1}) and (\ref{del_t_psi_1}) we can write $\partial_t v^r_1$ and $\partial_t \rho_1$ in terms of partial derivatives of $\Psi_1$ as
\begin{equation}\label{del_t_rho_1_and_v_1}
\begin{aligned}
& \partial_t v^r_1=\frac{1}{\sqrt{-\tilde{g}}H_0\rho_0 \tilde{\Lambda}}[(v^t_0 (1+\beta )+\rho_0\eta_1)\partial_t\Psi_1+(1+\beta )v^r_0\partial_r\Psi_1]\\
& \partial_t \rho_1=-\frac{1}{\sqrt{-\tilde{g}}H_0\rho_0 \tilde{\Lambda}}[\rho_0\eta_2\partial_t\Psi_1+\rho_0\partial_r\Psi_1]
\end{aligned}
\end{equation}
where $\tilde{\Lambda}$ is given by
\begin{equation}\label{Lambda_tilde}
\tilde{\Lambda}=(1+\beta )\left[\frac{g_{rr}(v^r_0)^2}{\Lambda v^t_0}-v^t_0\right]+\frac{c_{s0}^2}{\Lambda v^t_0}[\Lambda (v^t_0)^2-1-g_{rr}(v^r_0)^2].
\end{equation}

Now we first linearly perturb eq. (\ref{conserve}) and then take it's time derivative, which in turn gives
\begin{widetext}
\begin{equation}\label{w_mass_2}
\partial_t\left(h_0g_{rr}\partial_t v^r_1 \right)+\partial_t\left( \frac{h_0g_{rr}c_{s0}^2 v^r_0}{\rho_0}\partial_t \rho_1\right)-\partial_r\left( h_0\partial_t v_{t1}\right)-\partial_r\left( \frac{h_0 v_{t0}c_{s0}^2}{\rho_0}\partial_t \rho_1\right)=0 .
\end{equation}
\end{widetext}
We can write
\begin{equation}\label{del_t_pert_v_lower_t}
\partial_t v_{t1}=\tilde{\eta}_1\partial_t \rho_1+\tilde{\eta}_2 \partial_t v^r_1
\end{equation}
with
\begin{equation}\label{eta_2_tilde_and_eta_2_tilde}
\tilde{\eta}_1=-\left(\Lambda \eta_1+\frac{g_{\phi t}v_{\phi 0}c_{s0}^2}{g_{\phi\phi}\rho_0} \right),\quad\tilde{\eta}_2=-\Lambda \eta_2.
\end{equation}
Using eq. (\ref{del_t_pert_v_lower_t}) in the eq. (\ref{w_mass_2}) and dividing it by $h_0 v_{t0}$ yields
\begin{widetext}
\begin{equation}\label{w_mass_3}
\partial_t\left(\frac{g_{rr}}{v_{t0}}\partial_t v^r_1 \right)+\partial_t\left( \frac{g_{rr}c_s^2 v^r_0}{\rho_0 v_{t0}}\partial_t \rho_1\right)-\partial_r\left( \frac{\tilde{\eta}_2}{v_{t0}}\partial_t v^r_1\right)-\partial_r\left( (\frac{\tilde{\eta_1}}{v_{t0}}+\frac{c_s^2}{\rho_0})\partial_t \rho_1\right)=0
\end{equation}
\end{widetext}
where we use $h_0v_{t0}={\rm constant}$. Finally replacing $\partial_t v^r_1$ and $\partial_t \rho_1$ in eq. (\ref{w_mass_3}) using eq. (\ref{del_t_rho_1_and_v_1}) one obtains
\begin{widetext}
\begin{eqnarray}\label{w_mass_final}
\partial_t\left[k(r)\left(-g^{tt}+(v^t_0)^2(1-\frac{1+\beta}{c_s^2}) \right)\right]+\partial_t\left[ k(r)\left(v^r_0v_0^t(1-\frac{1+\beta}{c_s^2}) \right)\right] \nonumber \\
+\partial_r \left[ k(r)\left(v^r_0v_0^t(1-\frac{1+\beta}{c_s^2}) \right)\right]+\partial_r \left[ k(r)\left( g^{rr}+(v^r_0)^2(1-\frac{1+\beta}{c_s^2})\right)\right]=0
\end{eqnarray}
\end{widetext}

where $k(r)$ is a conformal factor whose exact form is not required for the present analysis.

Eq. (\ref{w_mass_final}) can be written as
\begin{equation}\label{wave-eq}
\partial_\mu (f^{\mu\nu}\partial_\nu \Psi_1)=0
\end{equation}
where $f^{\mu\nu}$ is obtained from the symmetric matrix
\begin{widetext}
\begin{eqnarray}\label{f_mass}
f^{\mu\nu}= k(r) \left[\begin{array}{cc}
-g^{tt}+(v^t_0)^2(1-\frac{1+\beta }{c_s^2}) & v^r_0v_0^t(1-\frac{1+\beta }{c_s^2})\\
v^r_0v_0^t(1-\frac{1+\beta }{c_s^2}) & g^{rr}+(v^r_0)^2(1-\frac{1+\beta }{c_s^2})
\end{array}\right]
\end{eqnarray}
\end{widetext}

The equation (\ref{wave-eq}) describes the propagation of the perturbation $ \Psi_1 $ in $1+1$ dimension effectively.\\
Eq. (\ref{wave-eq}) has the same form of a massless scalar field in curved spacetime (with metric $ g^{\mu\nu} $) given by
\begin{equation}\label{scalarfield}
\partial_\mu(\sqrt{-g}g^{\mu\nu}\partial_\nu \varphi)=0
\end{equation}
where $ g $ is the determinant of the metric $ g_{\mu\nu} $ and $\varphi$ is the scalar field. Comparing equation (\ref{wave-eq}) and (\ref{scalarfield}), the components of acoustic spacetime metric $ G_{\mu\nu} $ turns out to be
\begin{widetext}
\begin{equation}\label{Gmunu_NT}
G_{\mu\nu}^{NT} = k_1 (r) \begin{bmatrix}
-g^{rr}-(1-\frac{1+\beta}{c_{s0}^2})(v^r_0)^2 & v^r_0 v^t_0(1-\frac{1+\beta}{c_{s0}^2})  \\
v^r_0 v^t_0(1-\frac{1+\beta}{c_{s0}^2})  & g^{tt}-(1-\frac{1+\beta}{c_{s0}^2}) (v^t_0)^2
\end{bmatrix}
\end{equation}
\end{widetext}

where $ k_1 (r) $ is also a conformal factor arising due to the process of inverting $ G^{\mu\nu} $ in order to yield $ G_{\mu\nu} $. For our present purpose we do not need the exact expression for $ k_1 (r) $. The suffix `NT' is added because after the derivation of this acoustic metric, it is convenient to use just one $\beta$ and only flow with disc height as prescribed by Novikov and Thorne has non-zero $\beta$. Thus the above expression can now be used only to denote the acoustic metric for flow with height function as prescribed by Novikov and Thorne, whereas one can put $\beta = 0$ to get the acoustic metric for conical flow which is given by
\begin{widetext}
\begin{equation}\label{Gmunu_CF}
G_{\mu\nu}^{CF} = k_1 (r) \begin{bmatrix}
-g^{rr}-(1-\frac{1}{c_{s0}^2})(v^r_0)^2 & v^r_0 v^t_0(1-\frac{1}{c_{s0}^2})  \\
v^r_0 v^t_0(1-\frac{1}{c_{s0}^2})  & g^{tt}-(1-\frac{1}{c_{s0}^2}) (v^t_0)^2
\end{bmatrix}
\end{equation}
\end{widetext}

From stationary solution of accretion flow, one expects the sonic point to be the acoustic horizon of the Analogue metric. But the non-trivial structure of the metric corresponding to flow with disc height as prescribed by Novikov and Thorne defined in eq. (\ref{Gmunu_NT}) does not anymore assure that. Moreover, for both the metric, the particular coordinate assures that setting $G^{rr} = 0$ will determine the condition at acoustic black hole horizon. Thus we see that the condition at acoustic horizon for conical flow will be obtained by putting $\beta = 0$ in eq. (\ref{f_mass}). This yields
\begin{equation}
g^{rr}+(v^r_0)^2(1-\frac{1}{c_s^2}) = 0
\end{equation}
which in turn yields $u_0=c_{s0}$ at the acoustic horizon. But the condition at acoustic horizon for flow with height prescription as proposed by Novikov and Thorne will be
\begin{equation}
g^{rr}+(v^r_0)^2(1-\frac{1+\beta }{c_s^2}) = 0
\end{equation}
 which in turn yields
 \begin{equation}
 u_0 = c_{s0}/\sqrt{1 + \beta}
 \end{equation}
 at the acoustic horizon. For later purpose it will be convenient to define an effective sound speed
 \begin{equation}\label{effective_sound_speed}
 c_{eff} = c_{s0}/\sqrt{1 + \beta}
 \end{equation}
 for flow with height as prescribed by Novikov and Thorne and $c_{eff} = c_{s0}$ for conical flow.

\section{Carter Penrose Diagram of The Acoustic Metrics for Polytropic Accretion}

The redefinition of sound speed as performed in the previous section was motivated using the local conditions at black hole horizons. One may formally analyse the causal structure of the acoustic space-time of the models to find the location of the horizon. In this process the global features of light cones at any point in the acoustic sapcetime will then justify the redefinition of sound speed anywhere in the flow and not only at the acoustic black hole horizons. Thus in this section, we will study the causal structure by numerically plotting the Carter Penrose diagram \cite{wald1984} of the acoustic metric. The method of Carter Penrose diagram is used in general relativity to understand the global features of complicated black hole space times like Kerr spacetime and it is also used in cosmology \cite{Fre2}. Here we will use the Carter Penrose diagrams as it is used in the context of black hole space times to determine black hole and white hole regions. Carter Penrose Diagram of analogue metric has been done in \cite{Barcelo2004}, where constant sound speed was used and the flow profiles were assumed instead of being derived from some fluid or any other underlying set of governing equations. In our analysis, the sound speed is a local function of radial distance and the flow profile was solved from the governing equations. The non trivial discontinuous transition of the fluid profiles from supersonic region to subsonic region at the shock location is also the result of physical process determined by Rankine-Hugoniot conditions. This kind of complexities were not considered previously in the causal structure analysis of acoustic space times.

In this section we first simplify the metric elements in terms of stationary fluid variables and then apply proper coordinate transformation to remove the coordinate singularities of the metrics. Then those coordinates are further transformed to compactify the entire analogue spacetime. Next we plot the Carter Penrose diagram for both the models, define the boundaries from the point of view of differential geometry and then use causal relation to find out the features of different region of the analogue spacetime corresponding to different parts of the fluid flow. From this analysis the redefinition of sound speed will be justified globally.

\subsection{Acoustic Metric and Other Preliminaries}
From the obtained acoustic metric the line element is given by
\begin{equation}\label{line_element}
ds^2 = G_{tt} dt^2 + 2G_{tr} dt dr + G_{rr} dr^2 .
\end{equation}
where, the metric elements $G_{\mu \nu}$ are given by (\ref{Gmunu_CF}) and (\ref{Gmunu_NT})). The overall factor $k_1 (r)$ is not explicitly taken into account as we want to finally focus on conformally transformed metric where this factor can be absorbed with the conformal factor.

Using eq.(\ref{v^r}) and the contravariant form of eq. (\ref{v_t}), the metric elements of acoustic metrices turn out to be 

\begin{eqnarray}\label{metric_elements}
\left.\begin{aligned}
G_{tt} &=\frac{u_0^2-c_{\rm eff}^2}{c_{\rm eff}^2(1-u_0^2) g_{rr}}\\
G_{tr} &=G_{rt}=\frac{u_0(1-c_{\rm eff}^2)F_1 (r,\lambda)}{c_{\rm eff}^2(1-u_0^2)}\\
G_{rr} &=\frac{g_{rr} F_1^2 (r,\lambda )(1-c_{\rm eff}^2 )}{c_{\rm eff}^2(1-u_0^2)} - F_2 (r)
\end{aligned}\right.
\end{eqnarray}
where
\begin{eqnarray}\label{F_1 (r,lambda)}
\left.\begin{aligned}
F_1 (r,\lambda) &=\frac{g_{\phi \phi} +\lambda g_{\phi t}}{\sqrt{(g_{\phi\phi}+2\lambda g_{\phi t}-\lambda^2g_{tt})(g_{\phi \phi}g_{tt}+g_{\phi t}^2)g_{rr}}}\\
F_2 (r) &=\frac{g_{\phi \phi}}{g_{\phi \phi}g_{tt}+g_{\phi t}^2}
\end{aligned}\right.
\end{eqnarray}

Thus we see that at a critical point where $u_0^2=c_{\rm eff}^2$, the metric element $G_{tt}$ becomes zero. So we have to transform coordinate so that the co-ordinate singularity is removed. In the next section, we go through a systematic procedure such that the singularity is removed from new metric elements. In these new set of coordinates the spacetime represented by acoustic metric extends upto infinity. Thus we will conformally transform these coordinates such that the infinite spacetime is mapped into a finite region of some coordinate space. As mentioned in introduction, this conformal transformation and the corresponding Carter Penrose diagram will help to study the causal structure of the acoustic spacetime.

\subsection{Kruskal like Co-ordinate Transformation to Remove Singularity at Critical Points}\label{subsection:transformation}
The general transformations that lead to the construction of Carter Penrose diagrams in the context of black hole metrics has been studied in detail in \cite{wald1984, Fre2}. The corresponding transformations in the context of analogue metric has been derived in \cite{Barcelo2004}. Although we follow the general outline, the acoustic metric derived by us has the special feature of sound speed being a function of radial distance $r$. Thus the coordinate transformations are more involved and the procedure is described in this section in details.

First we choose null coordinates to write down the line element (\ref{line_element}). We note that for null or lightlike curves $ds^2 =0$, which yields
\begin{equation}
\left(dt-A_{+} (r) dr\right)\left(dt-A_{-} (r) dr\right)=0
\end{equation}
where,
\begin{equation}
A_{\pm} = \frac{-G_{tr} \pm \sqrt{G_{tr}^2-G_{rr}G_{tt}}}{G_{tt}} .
\end{equation}
So instead of co-ordinates $(t,r)$, we choose new co-ordinates to be null co-ordinates $(\chi ,\omega)$ such that
\begin{eqnarray}\label{null_co_ordinates_differential}
d\omega = dt -A_{+}(r)dr\\
d\chi = dt -A_{-}(r)dr
\end{eqnarray}
Using coordinate transformation introduced in (\ref{null_co_ordinates_differential}), the line element (\ref{line_element}) can be written as
\begin{equation}
ds^2 = G_{tt} d\chi d\omega
\end{equation}
After introducing the null coordinates, the next step usually should be the affine parametrization, which removes the removable singularity. But one can remove the singularities of $G_{tt}$ at the critical points by observing how the divergence behaves at the vicinity of horizon. To study this behaviour we expand $A_- (r)$ and $A_+ (r)$ upto first order of $(r-r_c )$. Thus by expanding $u_0$ near $r_c$ as
\begin{equation}\label{expansion}
u_0 (r) = -c_{\rm eff}(r_c )  + \left| \frac{du}{dr}\right|_{r_c }(r-r_c ) + O\left( (r-r_c)^2\right)
\end{equation}
where the negative sign of the effective sound speed implies the flow is towards the accretor. We also note that from(\ref{expansion}), near $r_c$ we can write
\begin{equation}\label{divergent_factor_upto_first_order}
u_0^2-c_{\rm eff}^2 \approx -2\left( c_{\rm eff}\frac{du}{dr}\right)_{r_c} (r-r_c )
\end{equation}
considering upto the first order term.\\
Now we expand $A_- (r)$ and $A_+ (r)$ up to linear order of $(r-r_c ) $. For that we first note that $G_{tt} \propto (u_0 - c_{eff})^2$ is very small near $r_c$ which implies $|G_{tt}G_{rr}/G_{tr}^2| \ll 1$. Thus we obtain
\begin{eqnarray}
A_+ (r) & = & \frac{ -G_{tr} + G_{tr}\left( 1- \frac{G_{tt}G_{rr}}{G_{tr}^2}\right)^{1/2}}{G_{tt} } \\
& \approx & -\frac{G_{rr}}{2G_{tr}}
\end{eqnarray}
and
\begin{eqnarray}
A_- (r) & = & \frac{ -G_{tr} - G_{tr}\left( 1- \frac{G_{tt}G_{rr}}{G_{tr}^2}\right)^{1/2}}{G_{tt} } \\
& \approx & -\frac{2G_{tr}}{G_{tt}}\\
& = & \frac{2F_{1}(r,\lambda)g_{rr}u_0 (c_{\rm eff}^2 -1)}{u_0^2 - c_{\rm eff}^2}\\
& \approx & \frac{F_{1}(r_c , \lambda)g_{rr}(u_{0c}^2 -1)}{u_{0c}' - (c_{eff}')_c}\frac{1}{r-r_c}\\
& = & \frac{1}{\kappa}\frac{1}{r-r_c}
\end{eqnarray}
where
\begin{equation}
\kappa = \frac{u_{0c}' - (c_{eff}')_c}{F_{1}(r_c , \lambda) g_{rr} (u_{0c}^2 -1)}.
\end{equation}
So, we see that although
\begin{equation}
\chi \approx t- \frac{1}{\kappa}\ln|r-r_c | 
\end{equation}
shows a logarithmic divergence at $r\rightarrow r_c$, the form of $G_{rr}$ and $G_{tr}$ ensure that
\begin{equation}
\omega = t + \int \frac{G_{rr}}{2G_{tr}} dr
\end{equation}
does not diverge at the critical points as the function inside the integral is regular there.

Thus we see that near the critical points
\begin{equation}\label{divergent_exponent}
e^{-\kappa \chi} \propto e^{-\kappa t}\left| r-r_c \right| \propto  e^{-\kappa t} (u_0^2-c_{\rm eff}^2)
\end{equation}
Now one can compare the acoustic null co-ordinates in this case with that of the Schwarzchild metric and guess a co-ordinate transformation such that the singularity of metric element at critical point is removed. The transformation equations can be given by
\begin{eqnarray}
\left.\begin{aligned}
U(\chi) &= -e^{-\kappa\chi}\\
W(\omega) &= e^{\kappa\omega}
\end{aligned}\right.
\end{eqnarray}
Using this new set of co-ordinates $(U,W)$, the line element can now be written as
\begin{equation}
ds^2 = G_{tt} \frac{e^{\kappa(\chi - \omega)} (u_0^2-c_{\rm eff}^2)}{\kappa^2 c_{\rm eff}^2(1-u_0^2){(1-2/r + a/r^2)^{-1}}}  dU dW .
\end{equation}
Thus at the numerator of the new metric element, the two factors multiplied together will cancel the divergence at critical points. Thus these new co-ordinates $(U, W)$ are similar to the Kruskal coordinates for the case of Schwarzchild metric which removes the co-ordinate singularity.

Now we have to compactify the infinite space into a finite patch of some coordinates. For that, the co-ordinates will be $(T, R)$ such that

\begin{eqnarray}
T= \tan^{-1} (W)+\tan^{-1} (U)\\
R= \tan^{-1} (W)-\tan^{-1} (U).
\end{eqnarray}

Just like one would plot $r=\rm{constant}$ lines in the Kruskal coordinate plane to get Kruskal diagram, $r=\rm{constant}$ lines can be drawn in these $(T,R)$ coordinates, and the resulting diagram would be able to represent causal structure of the original space time in a compactified region. The resulting diagram in $(T,R)$ coordinates is technically known as the Carter Penrose diagram.

Now we write down how the original metric defined in eq. (\ref{line_element}) is transformed when we express it in terms of $T$ and $R$ and comment on the structure of the metric. The metric $ds^2$, in terms of $T$ and $R$ is
\begin{equation}\label{conformal_metric}
ds^2 = \Omega^2 (-dT^2 + dR^2)
\end{equation}
where
\begin{widetext}
\begin{equation}
\omega^2 = -G_{tt}\sec^2(\frac{T+R}{2}) \sec^2(\frac{T-R}{2})  \frac{e^{\kappa(\chi - \omega)} (u_0^2-c_{\rm eff}^2)}{4 \kappa^2 c_{\rm eff}^2(1-u_0^2){(1-2/r + a/r^2)^{-1}}}.
\end{equation}\
\end{widetext}
We thus see from eq. (\ref{conformal_metric}) that we have finally obtained the transformations that connect the original metric conformally with two dimensional Minkwoski metric, where two metrics $G$ and $g$ on the same manifold $\mathscr{M}$ is said to be conformally connected if there is a positice definite conformal factor $\Omega^2 (x)$ such that
\begin{equation}\label{conformal_connection}
G_{\mu \nu} dx^{\mu} \otimes dx^{\nu} = \Omega^2 (x) g_{\mu \nu} dx^{\mu} \otimes dx^{\nu}.
\end{equation}
This conformal connection of the metric of analogue spacetime with two dimensional Minkowski spacetime will be the key point to establish causal significance of interesting features in analogue spacetimes presented earlier.
\subsection{Carter Penrose Diagram}
\begin{figure*}[!ht]
\begin{subfigure}{.5\textwidth}
    \centering
    \includegraphics[width=\columnwidth]{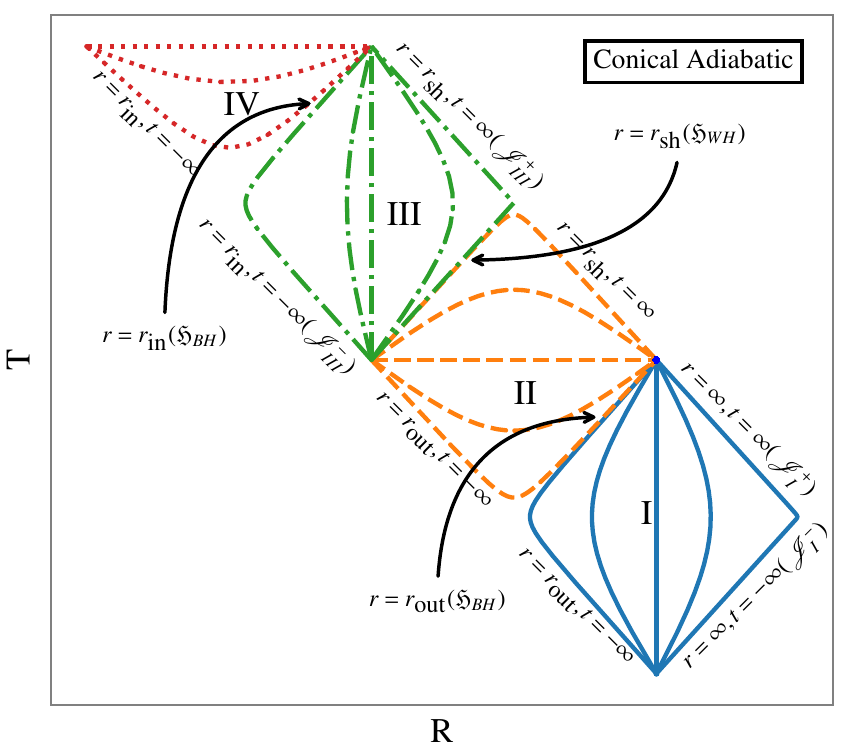}  
    \caption{Carter Penrose diagram for for adiabatic flow with conical disc height}
    \label{fig:conical_pcd_ad}
\end{subfigure}%
\begin{subfigure}{.5\textwidth}
    \centering
    \includegraphics[width=\columnwidth]{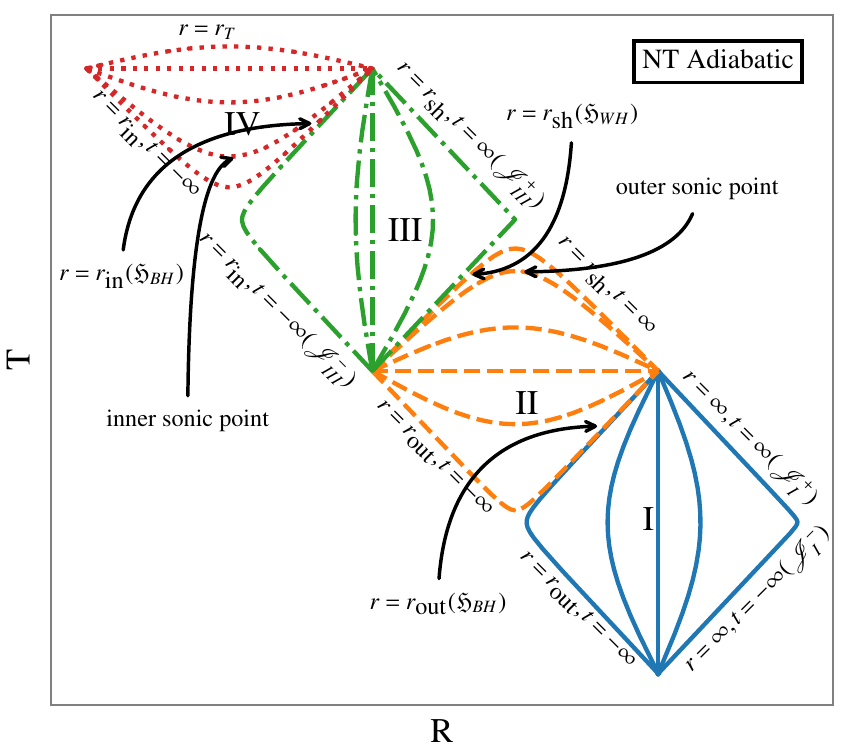}  
    \caption{Carter Penrose diagram for adiabatic flow with height expression formulated by Novikov and Thorne}
    \label{fig:NT_pcd_ad}
\end{subfigure}
\caption{In both the Carter Penrose diagrams for adiabatic flow, region $\rm I$ marked with blue solid lines corresponds to the flow outside outer critical point $r_{\rm out}$ up to infinity, where flow is subsonic. region $\rm II$ marked with yellow dashed lines corresponds to the flow inside $r_{\rm out}$ and outside $r_{\rm shock}$, where flow is supersonic. region $\rm III$ marked with green dashdotted lines corresponds to the flow inside shock $r_{\rm shock}$ and outside inner critical point $r_{\rm in}$, where flow is again subsonic. region $\rm IV$ marked with red dotted lines corresponds to the flow inside inner critical point $r_{\rm in}$ up to the minimum radius where flow can be extended outside the real Kerr black hole horizon and the flow is supersonic again. All the lines correspond to $r=\rm{constant}$ lines in their corresponding regions expect for the lines at boundaries where the $r=\rm{constant}$ lines coincide with $t=\rm{constant}$ lines where the time values are as specified in the figures. For flow with height functionas prescribed by Novikov and Thorne, the $r=\rm{constant}$ lines corresponding to usual sonic points are shown.}
\label{fig:pcd_ad}
\end{figure*}

In order to draw the Carter Penrose diagram, the multitransonic flow lines chosen previously are used as the background flow. Then the stationary metric elemnts are generated as a function of the the $r$ coordinate, i.e, the metric elements $G_{\mu\nu}$ are defined numerically on the analogue spacetime in  $(t,r)$ coordinates. Then by the prescribed transformations as formulated in the previous subsection, $r = $ constant lines are plotted in the $(T,R)$ plane, generating the Carter Penrose Diagram which transforms the infinite $(t,r)$ plane into a finite region of $(T,R)$ plane. The diagrams are obtained for adiabatic flow with both conical disc height and height function as prescribed by Novikov and Thorne. The corresponding plots are presented in fig. (\ref{fig:pcd_ad}) for both the geometries.

In order to understand the significance of the compactification which is presented in Carter Penrose diagram and to classify the regions corresponding to the black hole and the white hole, one must understand the causal significance the boundaries of the compactified diagrams. Once we define different parts of the boundaries, which already requires the understanding of causal relationship between two points in a spacetime, we will use the relationships on entire boundaries and see how the definitions of black hole and white hole regions arise naturally along with the corresponding horizons.

In order to understand why the compactification makes it easy to analyse causal structures from the compactified Carter Penrose diagrams, We must first state an lemma from the theory of differential manifolds that relates the  behaviour of null geodesics in analogue spacetime with the behaviour of null geodesics in Minkowski spacetime. The lemma states that \cite{Fre2}

\begin{lemma}
If two metrices $G$ and $g$ on the same manifold $\mathscr{M}$ are conformally related (as defined in eq. (\ref{conformal_connection})), then the null geodesics with respect to metric $G$ are null geodesics also with respect to the metric $g$ and vice-versa.
\end{lemma}

It was established in eq. (\ref{conformal_metric}) that the analogue metric derived in eq. (\ref{Gmunu_CF}) and eq. (\ref{Gmunu_NT}) for the multransonic flow chosen previously, is conformally flat, i.e, the resulting analogue metric is conformally connected with flat or Minkowski spacetime. Thus the sound cones in these compactified diagrams will have same orientation as light cones in Minkowski spacetime. So the sound cones in the compactified digrams are at an angle of $\pi /4$ and $3\pi/4$ to the horizontal lines if the scale of both $T$ and $R$ are the same. This orientation of sound cones is global.

The simple orientation of light cone in Carter Penrose diagram makes it easy to understand how timelike, spacelike and null vectors will be oriented, as they are just similar to the definition of vectors in two dimensional Minkowski spacetime. The vectors at any point in this diagrams will be timelike if they are inside the sound cone, null if they are on the boundaries of sound cone and spacelike if they are outside the sound cone.

With these definitions of the causal nature of vectors in mind we can now define causal curves in a manifold $\mathscr{M}$. A curve $\lambda(s)$ in any manifold $\mathscr{M}$ is a causal curve if for every point $p \in \lambda$ the tangent vector $t^\mu$ at that point is timelike or null. This definition of causal curve enables us to define the causal future/past of a point $p$. The causal future/past of a point $p$, denoted by $J^\pm (p)$ is a subset of $\mathscr{M}$ defined by the following condition
\begin{widetext}
\begin{equation}\label{causal_future_past}
J^\pm (p) = \left\{ q \in \mathscr{M} \mid \exists \textnormal{ future- (past-) directed causal curve } \lambda(s) \textnormal{ such that } \lambda(0) = p; \lambda(1) = q \right\}.
\end{equation}
\end{widetext}
The causal future/past of a region will be the union of causal future/past of all the points belonging to the region.

In the previous subsection (\ref{subsection:transformation}), we defined the set of transformations that transform the $(t,r)$ coordinates into $(T,R)$ coordinates. Thus we get an image of $(t,r)$ plane in the $(T,R)$ plane. Denoting this transformation or mapping $\psi$, we see that the mapping is injective but not surjective, i.e, the image of entire $(t,r)$ plane is a subset of $(T,R)$ plane, which represents the two dimensional Minkwoski plane as can be seen from the signatures in the metric.  Now the boundaries at infinity have been brought to finite distances and thus the boundary of this mapping in this finite domain can be analysed causally. We define the boundary of the mapping $\psi$ of the entire analogue spacetime $\mathscr{M}$ as
\begin{equation}
\partial \psi\left( \mathscr{M} \right) = i^0 \bigcup \mathscr{J}^+ \bigcup \mathscr{J}^-
\end{equation}
where
\begin{enumerate}
\item $i_0$, known as \textit{Spatial Infinity} is the endpoint of the $\psi$ image of all space-like curves in ($\mathscr{M}$, $g$).
    
\item $\mathscr{J}^+$, known as \textit{Future Causal Infinity} is the endpoint of the $\psi$ image of all future directed causal curves in ($\mathscr{M}$, $g$).
    
\item $\mathscr{J}^-$, known as \textit{Past Causal Infinity} is the endpoint of the $\psi$ image of all past directed causal curves in ($\mathscr{M}$, $g$).
\end{enumerate}

 In order to specify the boundaries of the Carter Penrose diagram as shown in fig. (\ref{fig:pcd_ad}), it must be noted that the whole diagram consists of four regions, denoted by \textrm{I}, \textrm{II}, \textrm{III} and \textrm{IV}. Region \textrm{I} and \textrm{III} are subsonic, keeping in mind that we do not anymore consider the usual sound speed as defined in eq. (\ref{cs-ad}) to seperate between subsonic region and supersonic region and consider the effective sound speed, by allowing the phonons to propagate freely in any direction. Thus both these regions have future and past causal infinities as their boundaries. These causal infinities are numerically generated by setting the time to be very large positive or negative numbers. The future null infinities for region \textrm{I} and region \textrm{III} are denoted by $\mathscr{J}_{\textrm{I}}^+$ and $\mathscr{J}_{\textrm{III}}^+$ respectively. Similarly, the past null infinities for region \textrm{I} and region \textrm{III} are denoted by $\mathscr{J}_{\textrm{I}}^-$ and $\mathscr{J}_{\textrm{III}}^-$ respectively.

Now the concepts of causal future and past are applied on the boundaries of the Carter Penrose diagrams to find out black hole and white hole regions formally. But before we delve into that an observation from the Carter Penrose diagrams should be noted. When we define a region as black hole or white hole in analogue spacetime, it is done with respect to either region \textrm{I} or region \textrm{III}. As previously mentioned, both regions \textrm{I} and \textrm{III} are subsonic, making them similar to regions of universe outside any kind of horizons. But as we see from the Carter Penrose diagrams, the two regions are connected by region \textrm{II}. As region \textrm{II} corresponds to supersonic flow, we can guess that this region will be black hole or white hole. As we will see next, this region can be denoted as a black hole or white hole region both, but it depends on the region with respect to which we define it's properties.

Now we focus on the features of region \textrm{II} and region \textrm{IV}. We find that the intersection of region \textrm{II} with the causal past of future null infinity of region \textrm{I} is null, i.e, $\textrm{II} \cap J^{-}(\mathscr{J}_{I}^+) = \emptyset$. Similarly we have $\textrm{IV} \cap J^{-}(\mathscr{J}_{III}^+) = \emptyset$. This property establishes a formal mathematical definition of a region which ican be defined as cut off from communication from the rest of the universe. As intuitively known, black hole region is characterised by this property of being cut off from the rest of the universe. Region \textrm{II} is cut off from region \textrm{I} and region \textrm{IV} is cut off from region \textrm{III}. Thus region \textrm{II} is a black hole region as perceived from an observer in region \textrm{I}, and region \textrm{IV} is a black hole region as perceived from region \textrm{III}. But as previously mentioned, in the context of analogue gravity, this property of the region \textrm{II} does not make it a black hole universally and it can only be termed black hole from the point of view of an observer in region \textrm{I}. But the Black hole horizons $\mathfrak{H}_{BH}$, the boundary of the black hole region separating it from the Causal Past of Future Null infinity, is a hypersurface whose definition does not require the specification of where the observer is located explicitly. Thus this particular causal boundary which is denoted as black hole horizon separates an black hole from its corresponding universe, or in terms of acoustic geometry with which we are concerned at this moment, the acoustic horizon is the barrier between a subsonic and supersonic region where we take the definition of effective sound speed as defined in eq. (\ref{effective_sound_speed}) into consideration while we define a region to be subsonic or supersonic. Mathematically, the definition of black hole horizon translates to
\begin{equation}
\mathfrak{H}_{BH} = \partial BH \cap \partial J^{-} \left( \mathscr{J}^{+} \right) .
\end{equation}
where $\partial$ denotes the boundary of the respective region.

From the definition of the black hole horizon, inspection of fig. (\ref{fig:pcd_ad}) implies that both the models, there are two black hole horizons for the individual model we chose. One of them is the boundary between region \textrm{I} and \textrm{II}, i.e, the outer critical point $r_{\rm out}$ and the other one is the boundary between region \textrm{III} and region \textrm{IV}, i.e, the inner critical point $r_{\rm in}$. This identification is true for accretion flow with conical height function as well as flow with height function as prescribed by Novikov and Thorne. In the case of Carter Penrose diagram for conical flow, the boundary of region \textrm{I} and region \textrm{II} denote the critical or sonic points which are the same in this case. But for flow  with height prescription as described by Novikov and Thorne, if the modified sound speed is not used, then we see that the line `$r = \textrm{outer sonic point}$' lies inside region \textrm{II} as has been pointed out. Thus, in this case, the sonic point does not act as the deciding boundary of propagation for linear perturbation. Thus the redefinition of effective sound speed is justified and one can use this definition not only at the horizon as motivated previously, but also this redefinition can be done in the entire acoustic spacetime or throughout the manifold where the flow is defined.

Now we focus on the causal significance of region \textrm{II} as observed from the point of view of an observer in region \textrm{III}. We find that the intersection of region \textrm{II} with the causal future of past null infinity of region \textrm{III} is null, i.e, $\textrm{II} \cap J^{+}(\mathscr{J}_{III}^+) = \emptyset$. This property establishes a formal mathematical definition of a region to which the universe can never communicate. But unlike the black hole, communication can be sent from the aforementioned region to its corresponding universe. Thus region \textrm{II} intuitively resembles the definition of a white hole in which no signal can be sent from the universe, i.e, region \textrm{III} but signals can reach to region \textrm{III} from region \textrm{II}. Here also the definition of white hole horizon $\mathfrak{H}_{WH}$ as the boundary of the white hole region separating it from the Causal Future of Past Null infinity does not require the specification of the position of an observer explicitly. Mathematically, the definition of white hole horizon translates to
\begin{equation}
\mathfrak{H}_{WH} = \partial WH \cap \partial J^{+} \left( \mathscr{J}^{-} \right) .
\end{equation}

From our discussion on the definition of black hole, we see, there is no event in the black hole region which causally affects any events in the corresponding universe. In the case of white hole, there is no event in the corresponding universe that will ever causally affect any event inside the white hole. It is evident that if the definition of modified sound speed is used then the critical points will act as the acoustic black hole horizon. The motivation for invoking the Carter Penrose diagram techniques to analogue spacetime was to establish the crucial role of critical points, but the role of shock as white whole horizon is not a feature that could be readily anticipated. In \cite{Abraham2006} Abraham et al. invoked the techniques of analogue gravity to establish that shocks act as white holes by invoking certain quantity which goes zero at the white hole horizon. Constant height of accretion disc in Kerr space time was used in this work. Here we conclude the same thing but the connection between Rankine-Hugoniot conditions and causal feature of the location of shock as established from our analysis with Carter Penrose diagram can further be studied.

\section{Phase Portrait and Carter Penrose Diagram for Isothermal Flow}
The problem of non isomorphism of critical points and sonic points for adiabatic flow with height function as prescribed by Novikov and Thorne was solved by invoking the definition of effective sound speed and establishing its global significance by causal structure analysis of Carter Penrose diagram. But we found that not only this solves the aforementioned nonisomorphism problem, but also establishes shock as white hole horizon. Thus the causal significance that can be extracted by analysing Carter Penrose Diagram is independent of the occurrence of any problem and can be pursued for its own merit to analyse analogue spacetime. Thus the previously used framework on the adiabatic flow can be used to analyse the analogue metric emerging from the isothermal flow also. In this section we present the phase portraits for isothermal flow and find out that the problem of non isomorphism of critical points and sonic points are not present for isothermal flow. After we present the stationary solutions and choose shocked multitransonic flow as before, the framework for Carter Penrose diagram is used and the corresponding diagrams are presented.

Now for isothermal flow, the governing fluid equations will have the same form as used in adiabatic flow. The characteristic of isothermal flow will be manifested in the results by the equation of state, given by,

\begin{equation}\label{eqtn_of_state_isothermal}
p=c_s^2\rho=\frac{\cal R}{\mu}\rho T=\frac{k_B\rho T}{\mu m_H}
\end{equation}

\noindent
where $T$ is the bulk ion temperature, $\cal R$ is the universal gas constant, $k_B$ is Boltzmann constant, $m_H$ is mass of the Hydrogen atom and $\mu$ is the mean molecular mass of fully ionized hydrogen. Now integral of motions must be constructed as was done for adiabatic flows.

\subsection{Phase Portrait}
The first conserved quantity obtained by integrating Euler equation (\ref{Euler}) turns out to be
\begin{equation}\label{conserved_isothermal}
\xi=v_t \rho^{c_s^2} = \rho^{c_s^2} \sqrt{\frac{\Delta}{B(1-u^2)}}.
\end{equation}

The second constant of integration obtained by integrating continuity equation has the same form as described for polytropic accretion, given by eq. (\ref{mass_accretion_rate}).

Now using these two constant of integration, we can obtain the derivative of advective velocity $u_0$ for isothermal flow as was done for polytropic flow. The expression for derivative of advective velocity corresponding to conical height function given by eq. (\ref{CFheight}) turns out to be (see \cite{Tarafdar_Deepika2019})
\begin{widetext}
\begin{equation}\label{dudr_CF_iso}
\frac{du_0}{dr}\Big|_{CF}^{iso} = \frac{u_0(1-u_0^2)\left[c_{s0}^2\frac{2r^2 - 3r + a^2}{\Delta r}+\frac{1}{2}(\frac{B'}{B}-\frac{\Delta'}{\Delta})\right]}{u_0^2- c_{s0^2}}=\frac{N_{CF}}{D_{CF}}|^{iso}.
\end{equation}
\end{widetext}
The expression of derivative for height function (see \cite{Tarafdar2021}) as prescribed by Novikov and Thorne given by eq. (\ref{NTheight}) is
\begin{widetext}
\begin{equation}\label{dudr_NT_iso}
\frac{du_0}{dr}\Big|_{NT}^{iso} = \frac{u_0(1-u_0^2)\left[c_{s0}^2(\frac{\Delta'}{2\Delta}+\frac{f'}{f})+\frac{1}{2}(\frac{B'}{B}-\frac{\Delta'}{\Delta})\right]}{u_0^2-c_{s0}^2}=\frac{N_{NT}}{D_{NT}}|^{iso}
\end{equation}
\end{widetext}
Now we see that for isothermal flow, one does not anymore have to worry about the issue of sonic points and critical points not being same as is evident from inspection of the denominator of both the derivatives stated above. Thus we will expect that Carter Penrose diagram will also establish this trivial feature and there is no need to differentiate between the sound speed and the speed of propagation of the first order perturbation in the accreting matter.

\begin{figure*}[!ht]
\begin{subfigure}{.5\textwidth}
    \includegraphics[width=.8\columnwidth]{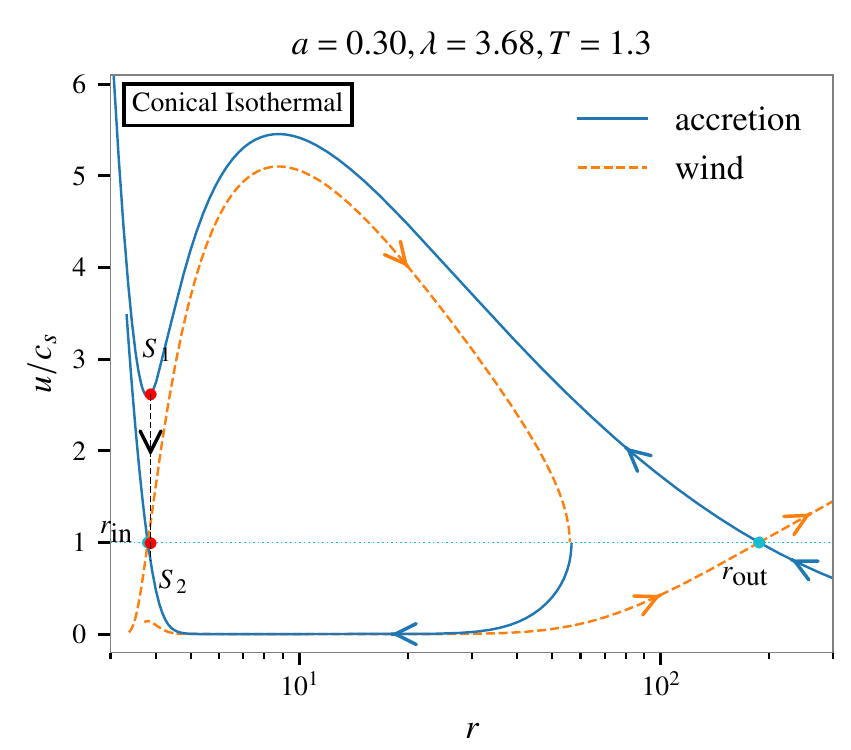}  
    \caption{Phase portrait for isothermal flow with conical disc height}
    \label{fig:conical_iso_phase}
\end{subfigure}%
\begin{subfigure}{.5\textwidth}
    \includegraphics[width=.8\columnwidth]{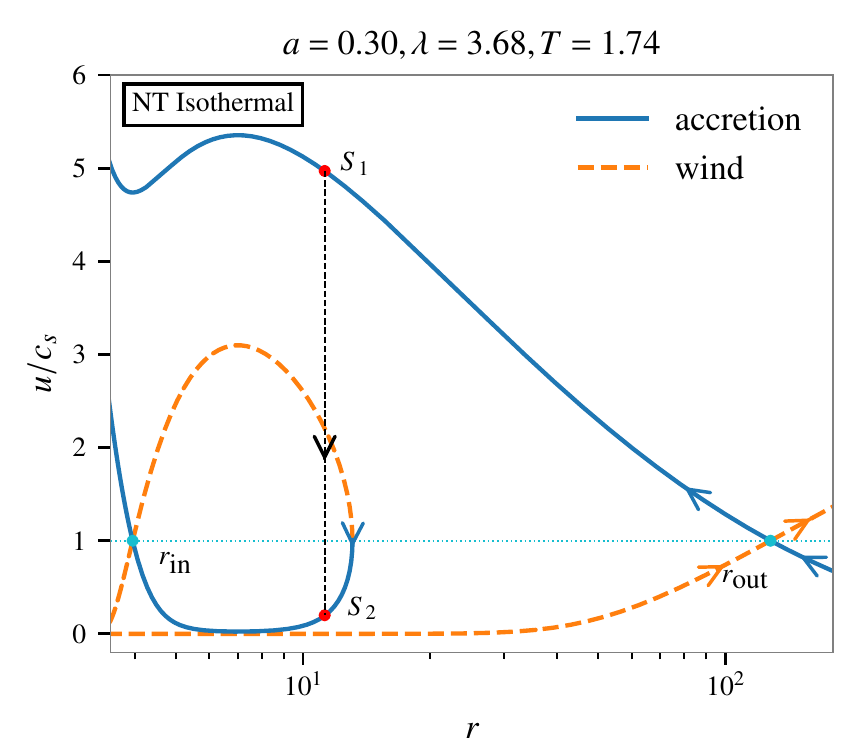}  
    \caption{Phase portrait for isothermal flow with height expression formulated by Novikov and Thorne}
    \label{fig:NT_iso_phase}
\end{subfigure}
\caption{Both phase portraits have been drawn for isothermal accretion with the set of parameter values $\lambda=3.68$ and $a=3.0$. The temperatures are written in the unit of $10^{10}$K. For both pictures the blue solid lines corresponds to accretion branch whereas the orange dashed lines correspond to wind branch. The innermost critical point is denoted as $r_{\rm in}$ and outermost critical point is denoted as $r_{\rm out}$. $S_1$ corresponds to the point in phase portrait in accretion branch through outer critical point where shock may occur and $S_2$ corresponds to the point where shock occurs in accretion branch through inner critical point. The same radial distance of $S_1$ and $S_2$ corresponds to the fact that the shock is infinitesimally thin. The dotted black line joining $S_1$ and $S_2$ corresponds to the discontinuous jump in the shock location. For both conical flow and flow with height as formulated by Novikov and Thorne, the sonic points and critical points coincide in the case of isothermal flow. The two points $r_{\rm in}$ and $S_2$ do not coincide in \ref{fig:conical_iso_phase}, although it may seem so from the figure as the radial distance between the two is very small.  }
 \label{fig:phase-portrait_iso}
\end{figure*}

We present the phase portraits for isothermal flow for both the conical disc height and flow  with height prescription as described by Novikov and Thorne in fig.(\ref{fig:phase-portrait_iso}).

From the family of phase portraits, only critical flows have been demonstrated in fig.(\ref{fig:phase-portrait_iso}). We will again choose the accretion flow from well outside the outer critical point, passing through outer critical point to shock, making  a discontinuous jump from the outer accretion branch to the inner accretion branch and then again becoming transonic at inner critical point and flowing inside it. The purpose of this is again to obtain transonic flow.

\subsection{Linear Perturbation scheme for Isothermal Flow}
The perturbation scheme will be same as used in polytropic flow and the time dependent accretion variables are again small time dependent linear perturbations added to the time independent stationary values as described in eq. (\ref{perturbations}).

\paragraph{Perturbation of Euler equation or the irrotationality condition} For isothermal flow, the irrotationality condition turns out to be \citep{Shaikh2017}

\begin{equation}\label{irrot_iso}
\partial_\mu(\rho^{c_s^2} v_\nu)-\partial_\nu(\rho^{c_s^2} v_\mu) = 0
\end{equation}
which can be obtained from the equation of state for isothermal flow along with the two fluid equations. From irrotationality condition (eq.(\ref{irrot_iso})). with $\mu=t$ and $\nu=\phi$ and with axial symmetry we have
\begin{equation}\label{irrotaionality_t_phi_iso}
\partial_t(h v_\phi)=0,
\end{equation}

\noindent
and, for $\mu=r$ and $\nu=\phi$ and the axial symmetry, we have

\begin{equation}\label{irrotationality_r_phi_iso}
\partial_r(\rho^{c_s^2} v_\phi)=0.
\end{equation}

\noindent
So $\rho^{c_s^2} v_\phi$ is a constant of motion and eq.(\ref{irrotaionality_t_phi_iso}) gives

\begin{equation}\label{del_t_v_phi_iso}
\partial_t v_\phi=-\frac{v_\phi c_s^2}{\rho}\partial_t \rho.
\end{equation}
which has exactly the same form as eq. (\ref{del_t_v_phi}), although in case of isothermal flow, $c_s$ is a constant whereas it was a function of radial distance in case of adiabatic flow. As eq. (\ref{del_t_v_up_phi}) to eq. (\ref{eta_1_eta_2_and_Lambda}) are derived from eq. (\ref{del_t_v_phi}), and they are not dependent on the geometry on the disc, rather on the background Kerr metric elements, these equations will remain the same for isothermal flow.

\paragraph{Perturbation of continuity equation} In case of isothermal flow for accretion disc in hydrostatic equilibrium along the vertical direction, i.e, disc with height function as prescribed by Novikov and Thorne, we have 
\begin{equation}
H(r) = \left(\frac{p}{\rho}\right)^{\frac{1}{2}} f(r) = c_s^2 f(r) = F(r)
\end{equation}
where $F(r)$ is purely a function of radial distance as sound speed $c_s$ is a constant in case of isothermal flow. In the case of conical flow, the height function is anyway a completely radial function where height does not depend on flow variable. Thus we do not need separate treatment for perturbation in the case for isothermal flow, as was necessary for adiabatic flows. Henceforth for isothermal case
\begin{equation}
H_{\theta 1} (r) = \frac{H_1 (r)}{r} = 0
\end{equation}
Thus the perturbed mass accretion rate here will have the form

\begin{equation}\label{Psi1_iso}
\Psi_1 = \sqrt{-g}[\rho_1 v_0^r H_{\theta 0}+\rho_0 v^r_1H_{\theta 0}]
\end{equation}
instead of eq. (\ref{Psi1}), which represented this perturbed quantity in case of adiabatic flow in hydrostatic equilibrium.

Using the definition of $\Psi$ and $\Psi_1$ from eq. (\ref{Sationary-mass-acc-rate}) and eq. (\ref{Psi1_iso}) in eq. (\ref{conserve}),one yields
\begin{equation}\label{del_r_psi_1_iso}
-\dfrac{\partial_r\Psi_1}{\Psi_0} = \dfrac{\eta_2}{v_0^r}\partial_t v^r_1+\dfrac{v_0^t}{v_0^r \rho_0 }\left[ 1 +\frac{\eta_1 \rho_0}{v_0^t}\right]\partial_t \rho_1,
\end{equation}
\noindent
and taking time derivative of eq. (\ref{Psi1_iso}), one yields
\begin{equation}\label{del_t_psi_1_iso}
\dfrac{\partial_t\Psi_1}{\Psi_0} = \dfrac{1}{v_0^r}\partial_t v^r_1+ \dfrac{\partial_t \rho_1}{\rho_0}
\end{equation}
instead of eq. (\ref{del_r_psi_1}) and eq. (\ref{del_t_psi_1}).

We see that eq. (\ref{del_r_psi_1_iso}) and eq. (\ref{del_t_psi_1_iso}) are basically eq. (\ref{del_r_psi_1}) and eq. (\ref{del_t_psi_1}) with $\beta = 0$. The reason for this is that there is no contribution of the first order perturbation of height function in the perturbation of mass accretion rate in eq. (\ref{Psi1_iso}) as was the case in eq. (\ref{Psi1}).

Thus eq. (\ref{del_t_rho_1_and_v_1}) and eq. (\ref{Lambda_tilde}) will be applicable for isothermal flow with $\beta = 0$.

Now putting $\mu = t$ and $\nu = r$ in the irrotatonality condition for isothermal flow, i.e, eq. (\ref{irrot_iso}),it is linearly perturbed and time derivative is taken. This yields
\begin{widetext}
\begin{equation}\label{w_mass_2_iso}
\partial_t\left(\rho_0^{c_s^2} g_{rr}\partial_t v^r_1 \right)+\partial_t\left( \frac{\rho_0^{c_s^2}g_{rr}c_{s0}^2 v^r_0}{\rho_0}\partial_t \rho_1\right)-\partial_r\left( \rho_0^{c_s^2}\partial_t v_{t1}\right)-\partial_r\left( \frac{\rho_0^{c_s^2} v_{t0}c_{s0}^2}{\rho_0}\partial_t \rho_1\right)=0.
\end{equation}
\end{widetext}
 which exactly resembles eq. (\ref{w_mass_2}), if $h_0$ in the aforementioned equation for adiabatic flow is replaced by $\rho_0^{c_s^2}$ for the isothermal case here. Now using eq. (\ref{del_t_pert_v_lower_t}) in eq. (\ref{w_mass_2_iso}), and dividing the equation by $\rho_0^{c_s^2}$  one yields eq. (\ref{w_mass_3}) again. Thus using $\partial_t v^r_1$ and $\partial_t \rho_1$ in eq.(\ref{w_mass_3}) using eq.(\ref{del_t_rho_1_and_v_1}) with $\beta = 0$ one obtains, 
 \begin{widetext}
  \begin{eqnarray}\label{w_mass_final_iso}
 \partial_t\left[ k(r)\left(-g^{tt}+(v^t_0)^2(1-\frac{1}{c_s^2}) \right)\partial_t \Psi_1\right]+\partial_t\left[ k(r)\left(v^r_0v_0^t(1-\frac{1}{c_s^2}) \right)\partial_r \Psi_1\right] \nonumber\\
 +\partial_r \left[ k(r)\left(v^r_0v_0^t(1-\frac{1}{c_s^2}) \right)\partial_t \Psi_1\right]+\partial_r \left[ k(r)\left( g^{rr}+(v^r_0)^2(1-\frac{1}{c_s^2})\right)\partial_r \Psi_1\right]=0
 \end{eqnarray}
\end{widetext}
 
 where $k(r)$ is a conformal factor whose exact form is not required for the present analysis, as mentioned in case of adiabatic flow. Eq.(\ref{w_mass_final}) can be written in the form of wave equation (\ref{wave-eq}).
 
 \noindent
 where in case of isothermal flow $f^{\mu\nu}$ is given by
 \begin{widetext}
 \begin{eqnarray}\label{f_mass_iso}
  f^{\mu\nu}= k(r) \left[\begin{array}{cc}
 -g^{tt}+(v^t_0)^2(1-\frac{1 }{c_s^2}) & v^r_0v_0^t(1-\frac{1}{c_s^2})\\
 v^r_0v_0^t(1-\frac{1}{c_s^2}) & g^{rr}+(v^r_0)^2(1-\frac{1}{c_s^2})
 \end{array}\right]
 \end{eqnarray}
 \end{widetext}
\noindent\\
The acoustic spacetime $ G_{\mu\nu} $ metric in the case of

isothermal flow thus turns out to be

\begin{widetext}
\begin{equation}\label{Gmunu_iso}
G_{\mu\nu} = k_1 (r) \begin{bmatrix}
-g^{rr}-(1-\frac{1}{c_{s0}^2})(v^r_0)^2 & v^r_0 v^t_0(1-\frac{1}{c_{s0}^2})  \\
 v^r_0 v^t_0(1-\frac{1}{c_{s0}^2})  & g^{tt}-(1-\frac{1}{c_{s0}^2}) (v^t_0)^2
\end{bmatrix}
\end{equation}
\end{widetext}
 
 \noindent
 where $ k_1 (r) $ is previously mentioned conformal factor arising due to the process of inverting $ G^{\mu\nu} $ in order to yield $ G_{\mu\nu} $. Again we do not need the exact expression for $ k_1 (r) $.
 
\subsection{Carter Penrose Diagram of Acoustic Metric for Isothermal Flow}
\begin{figure*}[!ht]
\begin{subfigure}{.5\textwidth}
    \centering
    \includegraphics[width=.8\columnwidth]{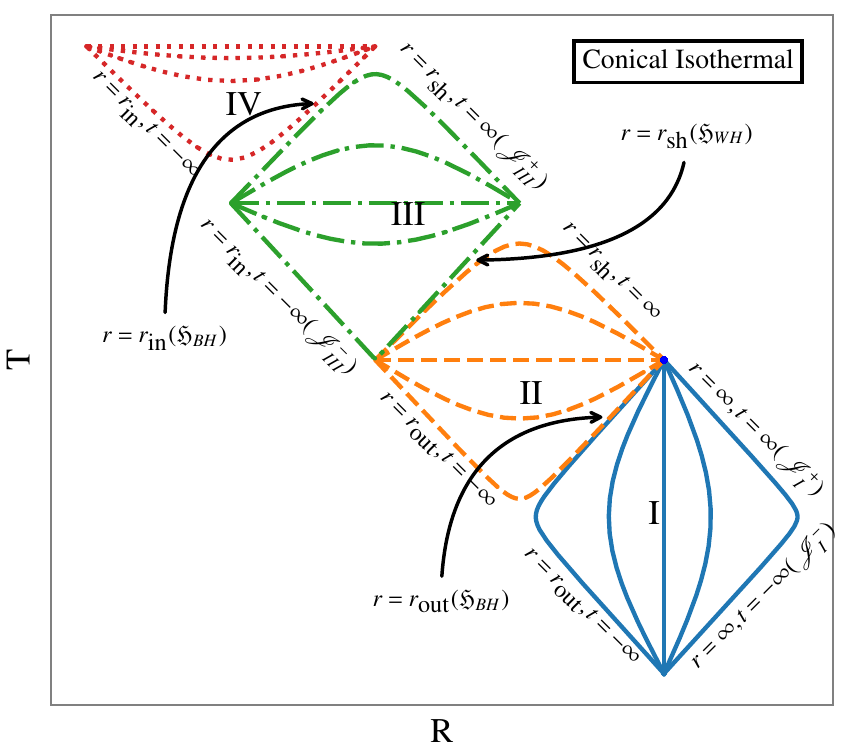}  
    \caption{Carter Penrose Diagram for isothermal flow with conical disc height}
    \label{fig:conical_iso}
\end{subfigure}%
\begin{subfigure}{.5\textwidth}
    \centering
    \includegraphics[width=.8\columnwidth]{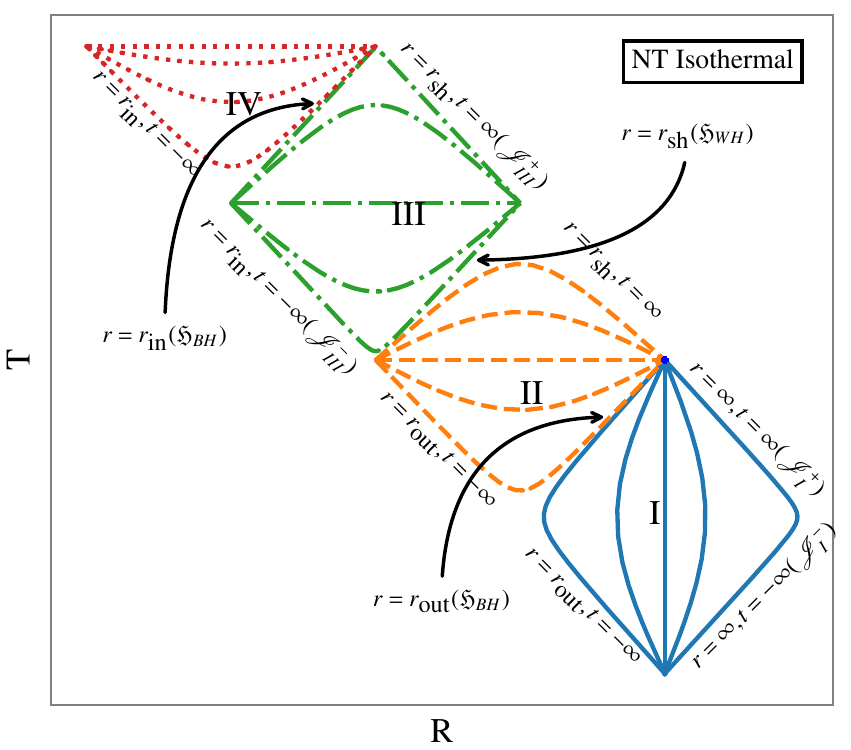}  
    \caption{Carter Penrose diagram for isothermal flow with height expression formulated by Novikov and Thorne}
    \label{fig:NT_iso}
\end{subfigure}
\caption{In both the Carter Penrose diagrams for isothermal flow, region $\rm I$ marked with blue solid lines corresponds to the flow outside outer critical point $r_{\rm out}$ up to infinity, where flow is subsonic. region $\rm II$ marked with yellow dashed lines corresponds to the flow inside $r_{\rm out}$ and outside $r_{\rm shock}$, where flow is supersonic. region $\rm III$ marked with green dashdotted lines corresponds to the flow inside shock $r_{\rm shock}$ and outside inner critical point $r_{\rm in}$, where flow is again subsonic. region $\rm IV$ marked with red dotted lines corresponds to the flow inside inner critical point $r_{\rm in}$ up to the minimum radius where flow can be extended outside the real Kerr black hole horizon and the flow is supersonic again. All the lines correspond to $r=\rm{constant}$ lines in their corresponding regions expect for the lines at boundaries where the $r=\rm{constant}$ lines coincide with $t=\rm{constant}$ lines where the time values are as specified in the figures.}
\label{fig:PCD_isothermal}
\end{figure*}
The construction of Carter Penrose diagrams are exactly the same as described for the case of adiabatic flow. All the coordinate transformations are same once one obtains the acoustic metric for a flow. The effective sound speed is same as the usual stationary thermodynamic definition of sound speed in case of isothermal accretion. The qualitative feature for the Penrose-Carter diagram is same as was illustrated in case of adiabatic flow. Here also multitransonicity gives rise to a pair of black holes connected by shock as illustrated in fig. \ref{fig:PCD_isothermal}. The similarity of the qualitative nature of the Carter Penrose diagrams invoke similar causal analysis as was done for adiabatic flow. The difference between the Carter Penrose diagram of adiabatic flow with height as prescribed by Novikov and Thorne and isothermal flows for both the disc heights is that whereas for the former case the sonic points are not the acoustic black hole horizons, for the later set of models, the sonic points and the acoustic black hole horizons are isomorphic. All the conclusions drawn in the case of adiabatic flow for conical disc height are although applicable for isothermal flows with both disc heights as the sonic points are isomorphic with the acoustic black hole horizons for these cases. Thus we restrain ourselves from the repetitive analysis and present the diagrams which are self explanatory following the discussion of adiabatic flow.

\section{Concluding remarks}
Accretion flow in hydrostatic equilibrium along the transverse direction differs from flows with other geometrical configuration in that, for flow in hydrostatic equilibrium the critical points do not coincide with the corresponding conventional sonic points. By conventional Mach number, we mean that the associated characteristic speed is assumed to be the stationary sound speed defined by $c_s^2=\gamma{p}/\rho$, whereas for acoustic geometry embedded within the accretion flow, the speed of propagation of the linear perturbation is taken to be the dynamical sound speed and is designated as the effective sound speed $c_{\rm eff}$ as defined at the critical point. Such stand alone role (in comparison to the other flow geometries) of the flow in hydrostatic equilibrium is because of the presence of the (stationary) sound speed in the expression of the disc height. The expression for the disc height is obtained by balancing the gravitational force with the pressure force and while doing so, certain set of idealized assumptions have been made where the derivative of the height function is approximated to the ratio of the local height to the radial distance in Newtonian limit. Such approximation is made since it has not been possible to construct and solve the Euler equation along the vertical direction (in addition to the radial Euler equation as defined and solved along the equatorial plane), and hence the radial sound speed enters in the expression of the disc height. A disc height obtained by such set of approximation is not completely realistic but that is the best one can do for accretion of ideal hydrodynamic fluid. For non ideal non hydrodynamic flow, certain other prescriptions are available which have been obtained by employing the non-LTE radiative transfer method or by using the Grad-Shafranov equations for the MHD flow \cite{Hubney1998, Davis2006, Beskin1997, Beskin2005, Beskin2010}. For our purpose, however, we stick to the ideal fluid for the shake of Lorentz invariance. 

We have demonstrated that an acoustic white hole forms at the shock location. At the shock location, the dynamical velocity as well as the characteristic sound speed changes discontinuously, and hence their space derivatives diverge. This does not allow us to compute the acoustic surface gravity at the shock location. Acoustic surface gravity at the white hole, thus, diverges. That is primarily because of the fact that the shock has been assumed to have infinitesimally small (practically zero) thickness. Had it been the case that we would consider a shock with finite thickness, the acoustic surface gravity at the acoustic white whole would be extremely large but finite. Possibility of having such unusually large acoustic surface gravity have been discussed in other context. It is to be mentioned that a shock with finite thickness may have different temperature at its two sides, leading to the dissipation of energy at the shock through radiation. Since our analogue model requires the fluid to be non-dissipative, we are somewhat compelled to consider shock with zero thickness only.

Fig. \ref{fig:PCD_isothermal} depicts the corresponding Carter Penrose diagram of isothermal accretion of matter flow in hydrostatic equilibrium along the transverse direction where the disc height is provided by the work of Novikov \& Thorne \cite{Novikov-Thorne1973}. It is evident from the figure that there is a lack of continuity between the region represented by yellow lines and that by the green lines. Such a trend in the causal structure indicates that the region from the outer sonic point to the shock location does not cover the entire manifold of analogue spacetime. The shape of the terminal green line representing the shock location proves that the shock is the boundary from which all future directed null sonic trajectories in the region between $r_{\rm out}$ and $r_{\rm sh}$  should escape ultimately to the subsonic region between $r_{\rm sh}$ and $r_{\rm in}$. This establishes shock to be the white hole with respect to any observer from the aforementioned subsonic region. The problem with this interpretation is that the region just outside the green line representing $r_{\rm sh}$ is not in the manifold, as mentioned. This is the artifact of the situation that the flow becomes discontinuous at the shock. This happens because the
kind of shock we consider has zero thickness, i.e, we have not been able to deal with a shock with finite thickness. It is expected that a continuous flow, which is a consequence of the finite width of the shock,  will not have the problem of exclusion of a part of the manifold.

For multitransonic shocked flow, accreting matter first encounters the outer sonic point, that is, as if, it `disappears' from the `outer acoustic universe' (space-time spanned from infinity/donor upto the outer sonic point) once it becomes supersonic for the first time. It means that once it crosses the outer acoustic black hole 
type horizon, i.e., the outer sonic point, no sonic signal emitted by any observer
co moving with the matter will be able to reach to any other observer situated in 
between infinity and the outer sonic point (observer comoving with subsonic flow). 
Once the supersonic matter, however, encounters the shock, the post shock accretion
flow resembles matter which has been `thrown out' to `another universe' through the
acoustic white hole, i.e., the stationary shock. The overall phenomena is equivalent to 
disappearance of matter from one universe through a black hole and re appearance of 
that matter to some other universe through a white hole. Such acoustic black hole 
white hole combination, thus, acts as a sonic analogue of wormhole. In our next 
work, we plan to introduce the properties of such acoustic worm holes in detail.

\bibliography{reference_susovan}

\end{document}